\documentclass[a4paper,preprint,12pt,sort&compress]{elsarticle}

\usepackage{mwe}
\usepackage{tikz}
\usepackage{structmech}
\usepackage{float}
\usepackage{caption}
\usepackage{subcaption}
\usepackage{amsmath}

\DeclareMathOperator*{\argmin}{arg\,min}
\usepackage{comment}
\usepackage{algorithm2e}
\SetKwComment{Comment}{/* }{ */}
\usepackage{amsthm}
\usepackage{amssymb}

\newtheorem{problem}{Problem}
\usepackage{expl3}
\ExplSyntaxOn
\newcommand\latinabbrev[1]{
  \peek_meaning:NTF . {
    #1\@}%
  { \peek_catcode:NTF a {
      #1.\@ }%
    {#1.\@}}}
\ExplSyntaxOff
\usepackage{mathrsfs}

\def\eg{\latinabbrev{e.g}}

\def\ie{\latinabbrev{i.e}}

\usepackage{amsmath,amssymb,amsthm,color,xfrac,mathrsfs}
\usepackage{bbm}
\usepackage{makecell}
\usepackage{epsfig,amsfonts}
\usepackage{graphicx,epsfig,epstopdf} 
\usepackage{array} 
\usepackage{color,soul}
\soulregister\citep{7}
\soulregister\citep{7}
\soulregister\ref{7}
\soulregister\pageref{7}
\sethlcolor{yellow}
\usepackage{appendix}
     
\usepackage{pmat} 
\usepackage{indentfirst} 
\usepackage{enumerate} 
\usepackage{lineno}
 \usepackage{placeins}
 \usepackage{silence}
\usepackage{booktabs,ctable,multirow,longtable} 
 \usepackage{rotating}
\usepackage{hyphenat}

\usepackage{changepage}

\NewDocumentCommand{\evalat}{sO{\big}mm}{%
  \IfBooleanTF{#1}
   {\mleft. #3 \mright|_{#4}}
   {#3#2|_{#4}}%
}

\DeclareMathAlphabet{\mathpzc}{OT1}{pzc}{m}{it}
\DeclareMathAlphabet{\mathcalligra}{OT1}{calligra}{m}{it}

\AtBeginDocument{
\let\div\undefined\DeclareMathOperator{\div}{{div}}
}

\newcolumntype{P}[1]{>{\centering\arraybackslash}m{#1}}

\graphicspath{{./}}

\newtheorem{remark}{Remark}
\journal{}
\makeatletter
\def\ps@pprintTitle{%
  \let\@oddhead\@empty
  \let\@evenhead\@empty
  \let\@oddfoot\@empty
  \let\@evenfoot\@oddfoot
}
\makeatother

\begin{document}
\begin{frontmatter}

\title{Phase-field modeling of brittle fracture in heterogeneous bars}
\author[1]{F. Vicentini}
\author[1]{P. Carrara}
\author[1]{L. De Lorenzis\corref{cor1}}
\ead{ldelorenzis@ethz.ch}

\address[1]{Eidgen\"{o}ssische Technische Hochschule Z\"{u}rich, Computational Mechanics Group, Tannenstrasse 3, 8092 Z\"{u}rich, Switzerland}
\cortext[cor1]{Corresponding author}


\begin{abstract}
We investigate phase-field modeling of brittle fracture in a one-dimensional bar featuring a continuous variation of elastic and/or fracture properties along its axis. Our main goal is to quantitatively assess how the heterogeneity in elastic and fracture material properties influences the observed behavior of the bar, as obtained from the phase-field modeling approach. The results clarify how the elastic limit stress, the peak stress and the fracture toughness of the heterogeneous bar relate to those of the reference homogeneous bar, and what are the parameters affecting these relationships. Overall, the effect of heterogeneity is shown to be strictly tied to the non-local nature of the phase-field regularization. Finally, we show that this non-locality may amend the ill-posedness of the sharp-crack problem in heterogeneous bars where multiple points compete as fracture locations.
\end{abstract}


\end{frontmatter}



\section{Introduction}
\label{sct:intro}
    
    Although at some scale all materials are heterogeneous, \ie~their properties vary in space, the adoption of homogeneous effective properties is often sufficient for mechanical modeling in engineering. However, the effect of heterogeneity cannot be ignored for a large class of mechanical problems such as those involving composite materials, biological tissues and metamaterials. The evolution of cracks in these materials follows complex patterns that challenge many modeling and computational approaches. 
    \par Phase-field modeling of brittle fracture was proposed by Bourdin et al. \citep{bourdin2000numerical} as the regularization of the variational fracture formulation by Francfort and Marigo \citep{francfort1998revisiting} and was later re-interpreted as a special family of gradient damage models \citep{pham2011gradient}. It provides a remarkably flexible variational framework to describe the nucleation and propagation of cracks with arbitrarily complex geometries and topologies in two and three dimensions \citep{ambati2015review}.   
     \par The original phase-field modeling approach is based on the assumption of homogeneous elastic and fracture properties of the material throughout the domain. Previous studies addressing phase-field modeling in heterogeneous materials adopt a pragmatic approach, by simply substituting the constant fracture toughness of the original model with a fracture toughness depending on the material point \citep{natarajan2019phase, kumar2021phase, hossain2014effective, shen2019novel}. Natarajan et al.~\citep{natarajan2019phase} propose a phase-field formulation for fracture in functionally graded materials. The approach is further developed by Kumar et al.~\citep{kumar2021phase}, where it is shown that the peak stress of a functionally graded material remains bounded between the values pertaining to the single constituents in homogeneous conditions. Hossain et al.~\citep{hossain2014effective} propose a technique based on phase-field modeling to evaluate the effective fracture toughness of heterogeneous media, while Shen et al.~\citep{shen2019novel} show that the introduction of a spatially variable fracture toughness in phase-field models is a promising tool to model fracture in bones. However, to the best of our knowledge the implications of heterogeneous material properties on the key predictions of the phase-field model have never been the object of a fundamental investigation. Thus, the relationship between local material properties and observed behavior as predicted by the phase-field model remains unclear, which in turn may prevent the proper calibration of the model and the proper interpretation of its results. 
    \par In this work, we perform such investigation for the one-dimensional case. We revisit the fundamental mathematical analysis in \citep{pham2011gradient} by assuming that the elastic and/or fracture material properties are heterogeneous with different possible profiles of spatial variations. We aim at quantitatively assessing how the heterogeneity in the material properties influences the observed behavior, and especially the peak stress and the fracture toughness, in the context of phase-field modeling. 
    \par The paper is structured as follows. Section~\ref{sct:PF1D} formulates the one-dimensional phase-field model of brittle fracture and the related evolution problem. Section 3 defines the classes and profile shapes of heterogeneity adopted in the subsequent sections. Section 4 briefly reviews the solution of the evolution problem for the homogeneous bar. The core of the study is Section~\ref{sct:het}, where the evolution problem is solved for the heterogeneous bar.  The analysis is first carried out in the one-dimensional space and then extended to a bar in the three-dimensional space. In Section~\ref{subsct:SvP}, we discuss the consequences of heterogeneity on the fracture behavior as predicted by the phase-field approach in contrast to the predictions of the sharp-crack model. The main conclusions are drawn in Section \ref{sct:concl}.
    \par In the following, the dependence on the pseudo-time $t$ of the quasi-static setting is denoted with the subscript $t$, \eg~$\alpha_t$ is the damage variable at pseudo-time $t$; the prime symbol denotes the derivative with respect to either the spatial coordinate $x$, \eg~$u_t'=\partial u_t/\partial x$, or the damage variable $\alpha$, \eg~$w'=dw/d\alpha$; the symbols $\nabla(\bullet)$ and $\Delta(\bullet)$ represent respectively the gradient and the Laplacian of the vectorial quantity $(\bullet)$ with respect to the spatial coordinates, while the divergence is indicated as $\div(\bullet)$; the dot symbol denotes the derivative with respect to the pseudo-time, \eg~$\dot{\alpha}_t=\partial \alpha_t/\partial t$; vectors and second-order tensors are denoted with bold symbols, \eg~$\boldsymbol{u}$ is the displacement vector in three-dimensions; the 2nd-order identity tensor is indicated with $\boldsymbol{I}$ and the symmetric part of the 4th-order identity tensor with $\mathbb{I}^s$; the superscript $T$ denotes the transpose of a matrix.


\section{One-dimensional phase-field model for brittle fracture}
\label{sct:PF1D}
In this section, we formulate the phase-field model of brittle fracture for a one-dimensional domain (Figure~\ref{fig:bar}), along the lines of \citep{pham2011gradient, pham2013onset,pham2010approche,pham2010approcheb} but with some generalizations to prepare for the later developments. The primary unknowns, both functions of the spatial coordinate $x$, are the \emph{displacement} $u$ and the \emph{phase-field} or \emph{damage variable} $\alpha$. The latter is an internal variable which describes the material damage level. Its magnitude is bounded between $\alpha=0$, corresponding to a sound material, and $\alpha=1$, denoting a fully damaged material.


\subsection{Energetic quantities}
As follows, we introduce some definitions which will be used in the remainder of this paper, especially concerning important energetic quantities. 
\label{subsct:en_q}
 The \emph{total energy density} $W$ is defined as
 \begin{equation}
     W(x,u',\alpha,\alpha'):=\varphi_{el}(x,u',\alpha)+\varphi_d(x,\alpha,\alpha')\text{,}
 \end{equation}
 where $\varphi_{el}$ is the \emph{elastic energy density} and $\varphi_{d}$ is the \emph{dissipation density}. The elastic energy density is given by
     \begin{equation}
         \varphi_{el}(x,u',\alpha):=\frac{1}{2}E_0(x)\,a(\alpha)\,u'^2\text{,}
     \end{equation}
where $E_0>0$ (considered here as a continuous function of $x$ to account for possible heterogeneity in the elastic properties of the material) is the \emph{undamaged elastic modulus} and $a(\alpha)$ is the \emph{degradation function}. The latter describes the degradation of the elastic modulus due to damage, thus it is a monotonically decreasing function such that $a(0)=1$ and $a(1)=a'(1)=0$. We also introduce the compliance modulation function $s(\alpha)$ as the reciprocal of $a(\alpha)$,
 \begin{equation}
    \label{eq:compl_mod_fun}
     s(\alpha):=\frac{1}{a(\alpha)}.
 \end{equation}
The dissipation density reads
     \begin{equation}
         \varphi_d(x,\alpha,\alpha'):=w_1(x)\,\left(w(\alpha)+\ell^2\,\alpha'^2\right)\text{,}
     \end{equation}
hence it consists of a local term, depending on the damage variable, and a non-local term, depending on its spatial derivative. In the local term, $w(\alpha)$ is the \emph{local dissipation function}, a monotonically increasing function of $\alpha$ such that $w(0)=0$ and $w(1)=1$. 
There are two common options for the definition of the functions $a(\alpha)$ and $w(\alpha)$ \citep{pham2011gradient,gerasimov2019penalization}:
         \begin{equation}
        \label{eq:AT1_dm}
        \texttt{AT1:}\hspace{3mm}a(\alpha)=(1-\alpha)^2 \hspace{3mm}\text{and}\hspace{3mm}w(\alpha)=\alpha\text{,}
        \end{equation}
        \begin{equation}
        \label{eq:AT2_dm}
        \texttt{AT2:}\hspace{3mm}a(\alpha)=(1-\alpha)^2 \hspace{3mm}\text{and}\hspace{3mm} w(\alpha)=\alpha^2\text{.}
        \end{equation}
Throughout this paper, we will focus on the \texttt{AT1} model.
In the non-local term, the dependency on the spatial derivative of the damage variable calls for the introduction of the internal length parameter $\ell$. The local magnitude of the dissipation density is modulated by a \emph{specific fracture energy} $w_1$, which is considered a continuous function of $x$ to account for possible heterogeneity in the fracture properties of the material.

The \emph{total energy functional} reads
 \begin{equation}
    \label{eq:tot_en_fun}
     \mathcal{E}(u,\alpha):=\int_{-L}^{L} W(x,u'(x),\alpha(x),\alpha'(x))\,dx
 \end{equation}
 and it is the sum of the \emph{elastic energy functional} $\mathcal{E}_{el}(u,\alpha)$ and of the \emph{dissipation functional} $\mathcal{D}(\alpha)$,
     \begin{equation}
        \mathcal{E}(u,\alpha)={\mathcal{E}_{el}}(u,\alpha)+{\cal{D}}(\alpha),
    \label{eq:energies}
    \end{equation}
    with
         \begin{equation}
        {\mathcal{E}_{el}}(u,\alpha):=\int_{-L}^{L} \varphi_{el}(x,u'(x),\alpha(x))\,dx=\int_{-L}^{L} \frac {1}{2}\,E_0(x)\,a(\alpha(x))\,u'(x)^2\,dx
    \label{eq:energies_el}
    \end{equation}
   and
            \begin{equation}
        \mathcal{D}(\alpha):=\int_{-L}^{L} \varphi_d(x,\alpha(x),\alpha'(x))dx=\int_{-L}^{L}w_1(x) \left(w(\alpha(x))+\ell^2\alpha'(x)^2\right)dx.
    \label{eq:energies_dam}
    \end{equation}

 
\subsection{Evolution problem}
 \label{subsct:evol}
\begin{figure}[H]
        \centering
            \begin{tikzpicture}[scale = 0.65]
		        
                \draw[line width=1 mm] (-4,0) -- (4,0);
		        
		        \draw [-latex] (0,0.5) -- (3, 0.5) node[midway, above,scale=1.]{$x$};
		        \draw (0,0) -- (0,0.5);
		        \node[scale = 1.] (origin) at (0,-0.5) {$0$};
		        
		        \draw [latex-latex] (-4,-1.5) -- (0, -1.5) node[midway, above, scale = 1.]{$L$};
		        \draw [latex-latex] (0,-1.5) -- (4, -1.5) node[midway, above, scale = 1.]{$L$};
		        
			    \draw[-latex] [ line width = 1] (4.2, 0) -- (5.5, 0)
			    node[midway, above, scale = 1.]{$U_t$};
		        
		        \draw[line width = 1] (-4,-1.8) -- (-4,1.8);
		        \foreach \nn in {-1.08, -0.36, 0.36, 1.08} {
			    \draw[line width = 1
			    ](-4,\nn) -- (-4.3,\nn - 0.3);
			    };
		        
		\end{tikzpicture}
        \caption{One-dimensional setting: clamped bar under tension.}
        \label{fig:bar}
    \end{figure}
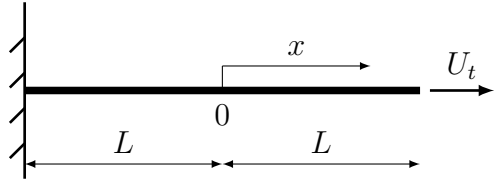

Let us now consider a bar clamped at the left end and loaded with a prescribed displacement at the opposite end (Figure~\ref{fig:bar}),
    
    \begin{equation}
        \label{eq:dirichlet}
        u_t\left(-L\right)=0
        \hspace{3mm}\text{and}\hspace{3mm}
        u_t\left(L\right)=U_t,
    \end{equation} 
    where $U_t$ is a positive smooth function of the pseudo-time $t$.
    
    \par At pseudo-time $t$, a displacement field $v$ and a damage field $\beta$ are admissible if they respectively belong to ${\cal{U}}_t$ and $\cal{A}$, with
    \begin{equation}
        {\cal{U}}_t:=\left\{v:v\left(-L\right)=0,v\left(L\right)=U_t\right\},
    \end{equation}
    \begin{equation}
        {\cal{A}}:=\left\{\beta: \beta\in[0,1]\right\}.
    \end{equation}
A more precise definition of these functional spaces is out of the scope of this work. We limit ourselves to require for them a sufficient regularity so that the total energy functional remains finite.
\par The evolution of the system can be studied as a quasi-static process parameterized through the pseudo-time $t\geq0$ and described with the function $t\mapsto(u_t,\alpha_t)$ and can be characterized variationally by means of total energy minimization. Following the variational approach in \citep{pham2013onset,pham2010approche,pham2010approcheb}, the evolution problem is governed by the principles of \emph{irreversibility}, \emph{local stability} and \emph{energy balance} and can be formulated as follows:
\begin{problem}[Evolution problem] Given the initial state $(u_0,\alpha_0)=(0,0)$ at the pseudo-time $t=0$, find $t\mapsto  (u_t,\alpha_t)\in \mathcal{U}_t\times\mathcal{A}$ fulfilling the following conditions:
\begin{enumerate}
  \item irreversibility: $t\mapsto\alpha_t$ is a non-decreasing function,
  \item local stability: 
    \begin{equation}
        \label{eq:loc_st}
        \begin{split}
            &\forall v \in {\cal{U}}_t,\forall \beta \in {\cal{A}}:\beta \geq \alpha_t,\exists \bar{h}>0 :\forall h\in
            [0,\bar{h}]\\
            &{\cal{E}}(u_t+h(v-u_t),\alpha_t+h(\beta-\alpha_t))\geq {\cal{E}}(u_t,\alpha_t),
        \end{split}
    \end{equation}
  \item  energy balance:      
    \begin{equation}
    \label{eq:energy_balance}
        {\cal{E}}(u_t,\alpha_t)={\cal{E}}(u_0,\alpha_0)+\mathcal{L}_t,
  \end{equation}
  where
\begin{equation}
    \label{eq:ext_work}
    \mathcal{L}_t:=\int_0^t \sigma_s(L)\,\dot{U}_s\,ds
\end{equation} 
is the work made by external actions in the pseudo-time interval $[0,t]$.
\end{enumerate}
\label{problem}
\end{problem}

In Eq.~\ref{eq:ext_work}, $\sigma_s$ denotes the Cauchy stress at pseudo-time $s$:
\begin{equation}
\label{eq:sigma_epsilon}
\sigma_s(x):=\evalat[\Bigg]{\frac{\partial W(x,u',\alpha_s(x),\alpha'_s(x))}{\partial u'}}{u'=u_s'(x)}=E_0(x)\,a(\alpha_s(x))\,u_s'(x).
\end{equation}


\subsection{First-order evolution problem} 

Upon a first-order expansion of Eq.~\ref{eq:loc_st} (under the assumption of sufficient smoothness), an evolution $t\mapsto (u_t,\alpha_t)$ is a solution of Problem~\ref{problem} \emph{only if} it is solution of the first-order evolution problem \citep{pham2013onset, pham2010approche, pham2010approcheb}
\begin{problem}[First-order evolution problem] Given the initial state $(u_0,\alpha_0)=(0,0)$ at the pseudo-time $t=0$, find $t\mapsto  (u_t,\alpha_t)\in \mathcal{U}_t\times\mathcal{A}$ sufficiently smooth fulfilling the following conditions:
\begin{enumerate}
    \item irreversibility:
        \begin{equation}
        \label{eq:ir}
            \dot{\alpha}_t\geq0,
        \end{equation}
    \item first-order stability:
        \begin{equation}
        \label{eq:st}
         \mathcal{E}'(u_t,\alpha_t)(v-u_t,\beta-\alpha_t)\geq0,\hspace{5mm}\forall (v,\beta)\in\mathcal{U}_t\times\mathcal{A}:\beta\geq\alpha_t,
        \end{equation}
    \item energy balance:
        \begin{equation}
        \label{eq:eb}
    \mathcal{E}'(u_t,\alpha_t)(\dot{u}_t,\dot{\alpha}_t)=\sigma_t(L)\,\dot{U}_t,
\end{equation}
\end{enumerate}
\label{problem2}
where $\mathcal{E}'(u_t,\alpha_t)(v,\beta)$ denotes the directional derivative of $\mathcal{E}$ at $(u_t,\alpha_t)$ in the direction $(v,\beta)$.
\end{problem}
Starting from Problem 2, standard arguments of Calculus of Variations deliver the equilibrium equation which states that the stress is constant along the bar \citep{pham2011gradient,pham2013onset, pham2010approche, pham2010approcheb}:
\begin{equation}
    \sigma'_t(x)=0\hspace{3mm}\text{in}\hspace{3mm}(-L,L),
    \label{eq:eq}
\end{equation}
as well as a set of Karush-Kuhn-Tucker (KKT) conditions
\begin{enumerate}
    \item\emph{irreversibility}: 
        \begin{equation}
        \label{eq:kkt_ir}
            \dot \alpha_t \geq 0\hspace{3mm}\text{in}\hspace{3mm}(-L,L),
        \end{equation}
    \item \emph{damage criterion}:
        \begin{equation}
            \label{eq:kkt_st}
            -\frac{1}{2}\,E_0\,a'(\alpha_t)\,u'^2_t \leq w_1\,w'(\alpha_t)
            -2\,w_1\,\ell^2\alpha_t''
            -2\,w_1'\,\ell^2\alpha'_t\hspace{3mm}\text{in}\hspace{3mm}(-L,L),
        \end{equation}
    \item \emph{loading-unloading conditions}: 
        \begin{equation}
        \label{eq:kkt_eb}
        \begin{aligned}
            \dot{\alpha}_t\left(\frac{1}{2}\,E_0\,a'(\alpha_t)\,u_t'^2+w_1\,w'(\alpha_t)-2\,w_1\,\ell^2\alpha_t''-2\,w_1'\,\ell^2\alpha_t'\right)=0\\
            \text{in}\hspace{3mm}(-L,L),
            \end{aligned}
        \end{equation}
  \end{enumerate}
along with the natural boundary conditions:
\begin{equation}
    \label{eq:natural1}
    \alpha_t'\left(-L\right)\leq 0,
        \hspace{5mm}
        \alpha_t'\left(L\right)\geq 0,
        \end{equation}
        \begin{equation}
        \label{eq:natural2}
        \alpha_t'\left(-L\right)\dot{\alpha}_t\left(-L\right) =0, 
        \hspace{5mm}
        \alpha_t'\left(L\right)\dot{\alpha}_t\left(L\right)=0.
\end{equation}

\begin{remark}
\label{differentiability}
    Eqs.~\ref{eq:eq}, \ref{eq:kkt_st}, \ref{eq:kkt_eb} involve the spatial derivative of $E_0(x)$ and $w_1(x)$ which do not appear in the classical homogeneous formulation.
\end{remark}

\begin{remark}  
The difference between the KKT conditions Eqs.~\ref{eq:kkt_ir}-\ref{eq:kkt_eb} for the general case of the heterogeneous bar problem and the analogous conditions for the special case of homogeneous bar is the presence of the terms containing the spatial derivative $w_1'$. This contribution to the strong form of the governing equations is not included in previous literature dealing with heterogeneous materials, see e.g. \textup{\citep{natarajan2019phase, kumar2021phase}}. However, these studies perform numerical finite element analyses based on the weak form associated to Eqs.~\ref{eq:kkt_ir}-\ref{eq:kkt_eb}; in the weak form the term containing $w_1'$ is compensated for by a similar term with opposite sign appearing after integration by parts, hence it does not influence results. As will be shown later, in the present study the same term is essential to understand the role played by heterogeneity on qualitative and quantitative aspects of the solution of the evolution problem.
\end{remark}


\section{Homogeneous and heterogeneous bars}
\label{sct:hom}
     Thus far,  the local elastic and fracture material properties have been characterized through $E_0(x)$ and $w_1(x)$, respectively. In the following, we distinguish between two cases:
 \begin{itemize}
     \item \emph{homogeneous bar}: the special case in which $E_0$ and $w_1$ are constant along the bar;
     \item \emph{heterogeneous bar}: the more general case in which $E_0$ and/or $w_1$ vary along the bar, as described by the functions $E_0(x)$ and/or $w_1(x)$, assumed to be sufficiently regular. 
 \end{itemize}
The spatial distribution of the material properties for the heterogeneous bar problem is further defined as
\begin{equation}
    \label{eq:het_def}
    E_0(x)=\bar{E}_0\cdot f_E(x)\hspace{3mm}\text{and}\hspace{3mm}w_1(x)=\bar{w}_1\cdot f_w(x),
\end{equation}
where the constants $\bar{E}_0$ and $\bar{w}_1$ are reference values of the undamaged elastic modulus and the specific fracture energy, respectively, and the functions $f_E(x)$ and $f_w(x)$ define the corresponding spatial variation profiles. The material with ${E}_0(x)=\bar{E}_0$ and $w_1(x)=\bar{w}_1$ is denoted as the homogeneous material \emph{associated} to a given heterogeneous material. In the following, all the quantities referred to the associated homogeneous material are denoted with a bar.
\par For the future developments, we now define three shapes of the heterogeneity profile, denoted by $h_i(x)$ with $i=lin,par,exp$, each of which depends on a length $\ell_f$ termed the \emph{characteristic length} of the heterogeneity (Table~\ref{tab:shapes}). The magnitude of $\ell_f$ characterizes how rapidly the material properties vary along the axis of the bar. This variation is associated to the slope of the profile ($i=lin$), to its curvature ($i=par$), or to both ($i=exp$).

We also define three classes of heterogeneous materials as summarized in Table~\ref{tab:classes}, each class assigning a profile shape to $f_w(x)$ and/or to $f_E(x)$. Accordingly, we distinguish between heterogeneity in the specific fracture energy (\texttt{hw}), heterogeneity in the undamaged elastic modulus (\texttt{hE}) and full heterogeneity (\texttt{hwE}).

\begin{remark}
We assume the material properties to be minimum at the midpoint cross-section of the bar and we choose symmetric increasing profiles of three different shapes (Table \ref{tab:shapes}). As a result, the midpoint cross-section is the weak location where we expect damage to start and develop first. We will compare the behavior of these heterogeneous bars with that of homogeneous bars where the properties are everywhere equal to the minimum values, i.e. $\bar E_0=\min_x E_0(x)$ and $\bar w_1=\min_x w_1(x)$.
\end{remark}

Finally, we introduce the dimensionless coordinate $\check{x}:=x/\ell$. With a slight abuse of notation, we denote $h_i$, $f_w$, $f_E$, $\alpha_t$ expressed as functions of $\check{x}$ with $h_i(\check{x})$, $f_w(\check{x})$, $f_E(\check{x})$, $\alpha_t(\check{x})$, respectively. In particular, the profile shapes are written in terms of $\check{x}$ as follows:
\begin{equation}
\label{eq:remapping}
    h_{lin}(\check{x})=1+r\,\lvert \check{x} \rvert\\ ,\;\;\;\;   h_{par}(\check{x})=1+r^2\, \check{x}^2\\  ,\;\;\;\;  h_{exp}(\check{x})=\text{exp}(r\, \lvert \check{x}\rvert)
\end{equation}
where $r$ is the \emph{characteristic ratio}:
\begin{equation}
    r:=\frac{\ell}{\ell_f}.
\end{equation}
The limit case $r\rightarrow 0$ is obtained for:
\begin{itemize}
    \item the sharp-crack model: $\ell\rightarrow 0$
    \item the associated homogeneous material: $\ell_f\rightarrow \infty$
\end{itemize}
Since we are interested in the diffusive approximation of cracks in heterogeneous materials, the relevant range of values for the applications we have in mind is $0<r<1$, i.e.~the variation of the material properties occurs over lengths that are sufficiently larger than the intrinsic length scale of the phase-field model. Values of $r$ (much) larger than 1 would represent a variation of material properties occurring at a length scale below the size of the characteristic length. Due to the nature of the regularization, such variations would not be “seen" by the model and are therefore not relevant for the present study.

\begin{table}[H]
	\begin{adjustwidth}{-3cm}{-3cm}
		\centering
	\begin{tabularx}{\textwidth}{ c  c  c } \toprule
	    
		\multirow{2}{*}{Shape} & 
		\multirow{2}{*}{Expression} &
		\multirow{2}{*}{Plot}\\ 
		\\\toprule
		
		linear  & $h_{lin}(x)=1+\frac{\lvert x \rvert}{\ell_f}$      & \raisebox{-.5\height}{\hspace*{-2.5mm}\includegraphics[scale=1]{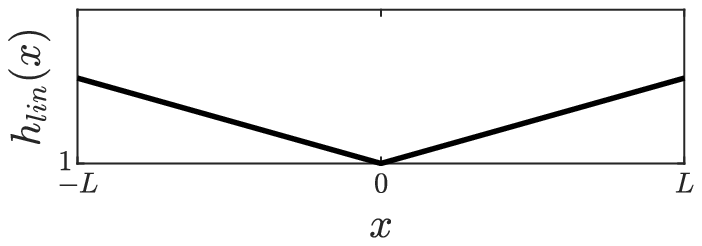}} \\
		parabolic  & $h_{par}(x)=1+\frac{x^2}{\ell_f^2}$ & \raisebox{-.5\height}{\hspace*{-2.5mm}\includegraphics[scale=1]{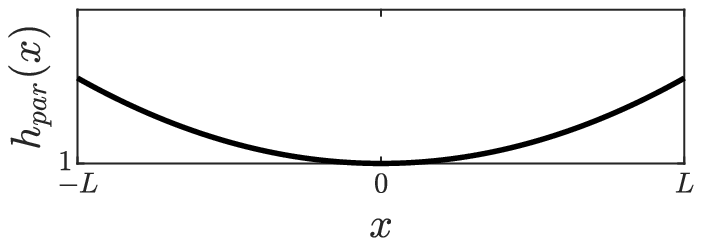}}      \\
		exponential & $h_{exp}(x)= \text{exp}\left(\frac{\lvert x \rvert}{\ell_f}\right)$ & \raisebox{-.5\height}{\hspace*{-2.5mm}\includegraphics[scale=1]{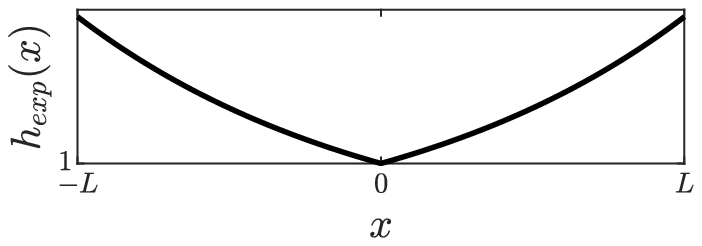}} \\
		
		\bottomrule
		\end{tabularx}
	\end{adjustwidth}
		\caption{Shapes of heterogeneity: linear shape, parabolic shape and exponential shape.} 
		 \label{tab:shapes}
	
	\end{table}

\begin{table}[H]
	\begin{adjustwidth}{-3cm}{-3cm}
		\centering
	\begin{tabular}{ccc} \toprule
	
		\multirow{2}{*}{Type} & 
		\multirow{2}{*}{$f_E(x)$} &
		\multirow{2}{*}{$f_w(x)$}\\ 
		\\\toprule
		
		\texttt{hw}  & 1      & $h_i(x)$ \\
		\texttt{hE}  & $h_i(x)$ & 1      \\
		\texttt{hwE} & $h_i(x)$ & $h_i(x)$ \\
		
		\bottomrule
        &\multirow{2}{*}{with $i=lin,par,exp$}
		\end{tabular}
	\end{adjustwidth}
		\caption{Classes of heterogeneity: heterogeneity in specific fracture energy (\texttt{hw}), heterogeneity in undamaged elastic modulus (\texttt{hE}), full heterogeneity (\texttt{hwE}).}
		 \label{tab:classes}
	\end{table}


\section{Solution of the evolution problem for the homogeneous bar}

In this section, we briefly summarize the solution of the evolution
problem formulated in Section \ref{sct:PF1D} for the special case of homogeneous
bar with undamaged elastic modulus $\bar{E}_{0}$ and specific fracture
energy $\bar{w}_{1}$. This solution has been thoroughly analyzed
in the literature, see e.g. \citep{pham2013onset, pham2010approche, pham2010approcheb}
. As mentioned earlier,
we limit ourselves to the case of the \texttt{AT1} model (Eq.~\ref{eq:AT1_dm}). This model
satisfies the \textit{strain hardening} condition, i.e.~$-w'\left(\alpha\right)/a'\left(\alpha\right)$
is increasing with respect to $\alpha$ (this implies that the elastic domain in the strain space expands for increasing damage), and the \textit{stress softening}
condition, i.e.~$w'\left(\alpha\right)/s'\left(\alpha\right)$ is
decreasing with respect to $\alpha$ for all values of $\alpha$  in $[0,1]$ (this implies that the elastic domain in the stress space shrinks for increasing damage) \citep{pham2011gradient}.

Let us consider an initially unstrained and undamaged bar, i.e.~$\left(u_{0},\alpha_{0}\right)=\left(0,0\right)$,
loaded with an imposed end displacement $U_{t}$ as introduced in
Section  \ref{subsct:evol}. The starting point of the analysis is the construction
of a \textit{homogeneous solution}, i.e.~a solution characterized
by a constant value of the damage variable along the bar under a monotonically increasing prescribed displacement. Since the
stress is constant due to equilibrium and the elastic properties are
homogeneous, the strain field is also constant and given by $u'_{t}=U_{t}/2L$.
The bar being initially undamaged, the solution of the evolution problem
is characterized by an initial \textit{elastic phase}, where the damage
criterion in Eq.~\ref{eq:kkt_st}, which in the case at hand simplifies to

\begin{equation}
-\frac{1}{2}\,\bar{E}_{0}\,a'\left(\alpha_{t}\right)u_{t}'^{2}\leq\bar{w}_{1}\,w'\left(\alpha_{t}\right),\label{eq:damcrit_simpl}
\end{equation}
is a strict inequality with $\alpha_{t}=0$. This phase continues
until the applied displacement $U_{t}$ reaches its value at the \textit{elastic
limit}

\begin{equation}
U_{e}=2L\sqrt{-\frac{2\,\bar{w}_{1}\,w'\left(0\right)}{\bar{E}_{0}\,a'\left(0\right)}}=2L\sqrt{\frac{\bar{w}_{1}}{\bar{E}_{0}}}
\end{equation}
corresponding to the \textit{elastic limit stress} or \textit{yield
stress}

\[
\sigma_{e}=\sqrt{\bar{E}_{0}\,\bar{w}_{1}}.
\]
We denote the corresponding pseudo-time as $t_{e}$. For $t>t_{e}$
(hence $U_{t}>U_{e}$), the damage criterion (\ref{eq:damcrit_simpl})
becomes an equality and damage can grow. For the ensuing homogeneous solution in the \emph{damaging phase} the homogeneous value of the damage variable can
be computed from the prescribed displacement $U_{t}$ through 

\begin{equation}
U_{t}=2L\sqrt{-\frac{2\bar{w}_{1}\,w'\left(\alpha_{t}\right)}{\bar{E}_{0}\,a'\left(\alpha_{t}\right)}}=2L\sqrt{\frac{\bar{w}_{1}}{\bar{E}_{0}\left(1-\alpha_{t}\right)}}\label{eq:hom_Ut}
\end{equation}
and the corresponding stress is

\begin{equation}
\sigma_{t}=\sqrt{\frac{2\,\bar{E}_{0}\,\bar{w}_{1}\,w'\left(\alpha_{t}\right)}{s'\left(\alpha_{t}\right)}}=\sqrt{\bar{E}_{0}\,\bar{w}_{1}\left(1-\alpha_{t}\right)^{3}}\label{eq:hom_sigmat}.
\end{equation}
Note that the validity of the strain hardening condition guarantees
that the functional relationship $\alpha_{t}\mapsto U_{t}$ in Eq.~\ref{eq:hom_Ut} is monotonically increasing, hence there is a unique
$\alpha_{t}$ solution for a given $U_{t}$, whereas the stress softening
property implies that the stress in Eq.~\ref{eq:hom_sigmat} decreases
with the damage level and hence with the applied displacement. Thus,
the \textit{peak stress} of the homogeneous response is reached for
$\alpha_{t}=0$, hence

\begin{equation}
\sigma_{p}=\sigma_{e}=\sqrt{\bar{E}_{0}\,\bar{w}_{1}}.
\end{equation}
Denoting the pseudo-time at peak stress as $t_{p}$, for the homogeneous
bar it is thus $t_{p}=t_{e}$. 

A stability analysis \citep{pham2013onset, pham2010approche, pham2010approcheb}
 demonstrates that for a sufficiently
long bar ($2L\gg l$) the homogeneous state is unstable for any $U_{t}\ge U_{e}$
and a damage localization necessarily arises at the end of the elastic
phase. In this \textit{localized solution}, damage is only non-zero
within an open interval $\mathcal{S}_{t}\in\left(-L,L\right)$ where
the damage criterion holds as an equality, while it vanishes in the
remainder of the domain. Infinite localized solutions are possible based on the position
of the damage localization region within the domain. Without loss of
generality, we assume here that this region is centered at $x^{*}=0$,
i.e. at the midpoint of the bar. A thorough analysis of the localization
phase can be found in \citep{pham2013onset, pham2010approche, pham2010approcheb}
 and references therein and is
not repeated here. During localization, the maximum value of the damage
variable, i.e.~$\alpha_{t}\left(x^{*}\right)$, increases monotonically
whereas the stress decreases, hence $\sigma_{p}$ is the peak stress
not only of the homogeneous response but of the overall stress-displacement
response. To follow this phase, control is switched from increasing prescribed displacement to decreasing stress or increasing $\alpha_{t}\left(x^{*}\right)$, as the corresponding $U_t$ may no longer be monotonically increasing depending on the length of the bar. 

At the end of the localization phase, $\alpha_{t}\left(x^{*}\right)$
reaches the value $1$ leading to failure of the bar. We denote the
corresponding pseudo-time as $t_{u}$, at which the stress $\sigma_{u}=0$
and the fully localized damage profile $\alpha_{u}$, symmetric about
$x^{*}=0$, reads  \citep{gerasimov2019penalization}:
\begin{align}
  \alpha_u(x)=
  \begin{cases}
    \hfil 0,                    
    & \text{in}\ [-L,-\delta_u] \\
    \frac{1}{4}\frac{x^2}{\ell^2}+\frac{x}{\ell}+1,
    & \text{in}\  \left(-\delta_u,0\right)
\end{cases},
\end{align}
where $\delta_u=2\ell$ is the half-support width.

The fracture toughness is defined as the dissipated energy at failure:

\begin{equation}
\label{G_c}
G_{c}:=\mathcal{D}\left(\alpha_{u}\right)
\end{equation}
and in the present case is given by

\begin{equation}
\label{eq:Gc_hom}
G_{c}=\frac{8}{3}\,\bar{w}_{1}\,\ell.
\end{equation}

\begin{remark}
    According to Eq.~\ref{eq:Gc_hom}, for a given $\ell$, knowledge of the local quantity $\bar{w}_1$ is sufficient to determine the global quantity $G_c$.
\end{remark}




\section{Solution of the evolution problem for the heterogeneous bar}
\label{sct:het}

In this section, we derive the solution of the evolution problem for the heterogeneous bar, including the homogeneous and the localized solutions, and especially focusing on the effect of the heterogeneity on peak stress and fracture toughness. As in the case of the homogeneous bar, we carry out the analysis for the \texttt{AT1} model.

\subsection{Governing equations on half domain}
\label{subsct:symmetry}

We can take advantage of symmetry and study the problem on half (e.g. on the left half) of the domain. For later reference, we rewrite here the equilibrium equation

\begin{equation}
    \sigma'_t(x)=0\hspace{3mm}\text{in}\hspace{3mm}(-L,0)
    \label{eq:eqs}
\end{equation}
and the KKT conditions
\begin{enumerate}
    \item\emph{irreversibility}: 
        \begin{equation}
        \label{eq:kkt_irs}
            \dot \alpha_t \geq 0\hspace{3mm}\text{in}\hspace{3mm}(-L,0),
        \end{equation}
    \item \emph{damage criterion}:
        \begin{equation}
            \label{eq:kkt_sts}
            -\frac{1}{2}\,E_0\,a'(\alpha_t)\,u'^2_t \leq w_1\,w'(\alpha_t)
            -2\,w_1\,\ell^2\alpha_t''
            -2\,w_1'\,\ell^2\alpha'_t\hspace{3mm}\text{in}\hspace{3mm}(-L,0),
        \end{equation}
    \item \emph{loading-unloading conditions}: 
        \begin{equation}
        \label{eq:kkt_ebs}
        \begin{aligned}
            \dot{\alpha}_t\left(\frac{1}{2}\,E_0\,a'(\alpha_t)\,u_t'^2+w_1\,w'(\alpha_t)-2\,w_1\,\ell^2\alpha_t''-2\,w_1'\,\ell^2\alpha_t'\right)=0\\
            \text{in}\hspace{3mm}(-L,0).
            \end{aligned}
        \end{equation}
  \end{enumerate}
 The governing equations above do not require the existence of $f_w'(0)$ but only of the left-hand derivative. The natural boundary conditions read:
\begin{equation}
    \label{eq:natural1s}
    \alpha_t'\left(-L\right)\leq 0,
        \end{equation}
        \begin{equation}
        \label{eq:natural2s}
        \alpha_t'\left(-L\right)\dot{\alpha}_t\left(-L\right) =0.
\end{equation}
\par We now need additional boundary conditions at $\check{x}=0$. These can be easily retrieved from the variational approach as for the case of the homogeneous bar \citep{pham2013onset}, and read as follows depending on $t$:

\begin{equation}
    \label{eq:bcs1}
    \alpha_t'(0)=0\hspace{3mm}\text{for}\hspace{3mm}t\in(t_e,t_u),
\end{equation}
\begin{equation}
    \label{eq:bcs2}
    \alpha_u(0)=1.
\end{equation}


\subsection{Homogeneous solution}
\label{subsct:dimless_p}
\par Using Eqs.~\ref{eq:AT1_dm} and \ref{eq:sigma_epsilon}, the damage criterion Eq.~\ref{eq:kkt_sts} takes the form
\begin{equation}
  \label{eq:damage_criterion_stress}
  \frac{1}{E_0}\,\frac{1}{(1-\alpha_t)^3}\,\sigma_t^2\leq w_1(1-2\,\ell^2\,\alpha_t'')-2\,w_1'\,\ell^2\alpha'_t\hspace{3mm}\text{in}\hspace{3mm}(-L,0).
\end{equation}
With the dimensionless coordinate defined in Section \ref{sct:hom}, the dimensionless damage criterion reads
\begin{equation}
    \label{eq:dimensionless_damage_criterion2}
    \begin{aligned}
  \frac{1}{f_E\left(\check{x}\right)}\,\frac{1}{(1-\alpha_t(\check{x}))^3}\,\check{\sigma}_t^2
  \leq
  f_w(\check{x})\left(1-2\,\alpha_t''(\check{x})\right)-2\,f_w'(\check{x})\,\alpha_t'(\check{x})\\
  \text{in}\hspace{3mm}\left(-L/\ell,0\right),
  \end{aligned}
\end{equation}
where $\check{\sigma}_t$ is the dimensionless stress
\begin{equation}
    \label{eq:rho_def}
    \check{\sigma}_t:=\frac{\sigma_t}{\bar{\sigma}_e}
\end{equation}
and $\bar{\sigma}_e=\bar{\sigma}_p=\sqrt{\bar{E}_0\,\bar{w}_1}$ is the yield stress, equal to the peak stress, for the associated homogeneous material.

As in the analysis for the homogeneous bar, we first look for a \emph{homogeneous solution} in the \emph{elastic phase}, where the damage criterion is satisfied as a strict inequality with $\alpha_t=0$. It is straightforward to determine the dimensionless elastic limit stress $\check{\sigma}_e$ and the position $\check{x}^*$ of the first point of the bar reaching the elastic limit as follows
\begin{equation}
    \label{eq:het_sigma_e}
    \check{\sigma}_e=\min_{\check{x}\in[-L/\ell,0]} \sqrt{f_E(\check{x})\cdot f_w(\check{x})}=1,
\end{equation}
\begin{equation}
    \check{x}^*=\argmin_{\check{x}\in[-L/\ell,0]} \sqrt{f_E(\check{x})\cdot f_w(\check{x})}=0.
\end{equation}
\begin{remark}
    According to Eq.~\ref{eq:het_sigma_e}, the elastic limit stress for the heterogeneous bar is the same as for the bar made of the associated homogeneous material and is equal to 
    \begin{equation}
    \label{eq:sigma_e}
    \sigma_e=\bar{\sigma}_e=\sqrt{\bar{E}_0\,\bar{w}_1}.
    \end{equation}
\end{remark}

Next, we look for a homogeneous solution in the \emph{damaging phase}, where the damage criterion is satisfied as an equality with $\alpha_t\neq0$ and uniform along the bar. Assuming uniform damage delivers \begin{equation}
  \label{eq:dc_S}
    \begin{aligned}
  \frac{1}{f_E\left(\check{x}\right)}\,\frac{1}{(1-\alpha_t)^3}\,\check{\sigma}_t^2
  \leq
  f_w(\check{x}) 
 \;\;\; \text{in}\hspace{3mm}\left(-L/\ell,0\right),
  \end{aligned}
\end{equation}
Eq.~\ref{eq:dc_S} only admits a solution for the special case where $f_E(x)\cdot f_w(x)$ is constant along the bar, which is excluded a priori by our choice of the heterogeneity profiles in Section \ref{sct:hom}. Hence, the evolution problem for the general case of a heterogeneous bar does not admit a homogeneous solution in the damaging phase.


\subsection{Localized solution}
\label{subsct:localized_sol}
After reaching the elastic limit, i.e.~for $t>t_e$, the heterogeneous bar problem admits only a localized solution with $\alpha_t(\check{x})\neq0$. Hence, within the left half-domain there exists an interval $(-\check{\delta}_t,0)$ where the damage criterion holds as an equality while the remainder of the bar is undamaged, i.e.
\begin{equation}
    \label{eq:dimensionless_damage_criterion}
    \begin{gathered}
  \frac{1}{f_E}\frac{1}{(1-\alpha)^3}\,\check{\sigma}_t^2
  =
  f_w\cdot \left(1-2\,\alpha_t''\right)-2\,f_w'\,\alpha_t'\hspace{3mm} \text{in}\hspace{3mm}\left(-\check{\delta}_t,0\right),\\\alpha_t=0\hspace{3mm}\text{in} \hspace{3mm}\left[-L/\ell,-\check{\delta}_{t}\right]
  \end{gathered}
\end{equation}
with $\check{\delta}_t:=\delta_t/\ell$, where $\delta_t$ is the half-support width (a priori unknown) at pseudo-time $t$. 
The regularity of the functions $a(\alpha)$ and $w(\alpha)$ implies that $\alpha_t$ and $\alpha_t'$ are continuous within $(-L/\ell,0)$ \citep{pham2013onset}, hence
\begin{equation}
    \label{eq:bcs_delta}
    \alpha_t(-\check{\delta}_t)=\alpha'_t(-\check{\delta}_t)=0,
\end{equation}
while the boundary conditions Eqs.~\ref{eq:bcs1}, \ref{eq:bcs2} continue to hold.

The damage profile $\alpha_t(\check{x})$ is assumed to be monotonically increasing over $(-\check{\delta}_t,0)$ (an assumption that can be easily verified a posteriori), therefore the maximum damage value is the value of the damage variable at $\check{x}^*=0$, \ie~$\alpha_t(0)$.

\subsubsection{Peak stress and stress-displacement curve during localization} 
\label{subsct:peak}
Next, we show that the boundary value problem in $\alpha_t$ constituted by Eq.~\ref{eq:dimensionless_damage_criterion} along with the boundary conditions in Eqs.~\ref{eq:bcs_delta}, \ref{eq:bcs1} for $t\in[t_e,t_u)$ admits solutions for increasing values of the stress, up to a peak value that, as in the case of the homogeneous bar, we denote as the \emph{peak stress}. After the peak, the stress decreases. As follows, we devise a semi-analytical scheme to solve the problem and determine the peak stress. 

\begin{figure}[H]
  \centering
     {\includegraphics{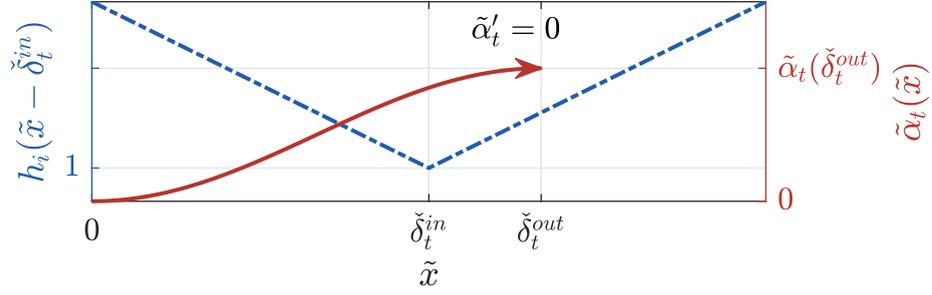}}
  \caption{Schematic representation of the procedure that returns a value  $\check{\delta}_t^{out}$ for any $\check{\delta}_t^{in}$. The blue and red lines represent the profile shape $h_i$ and the damage variable $\alpha_t$, respectively. Numerical integration starts from $\tilde{x}=0$ and is stopped at $\tilde{x}=\check{\delta}_t^{out}$.}
  \label{fig:D_function}
  \end{figure}

\begin{figure}[H]
  \centering
     {\includegraphics{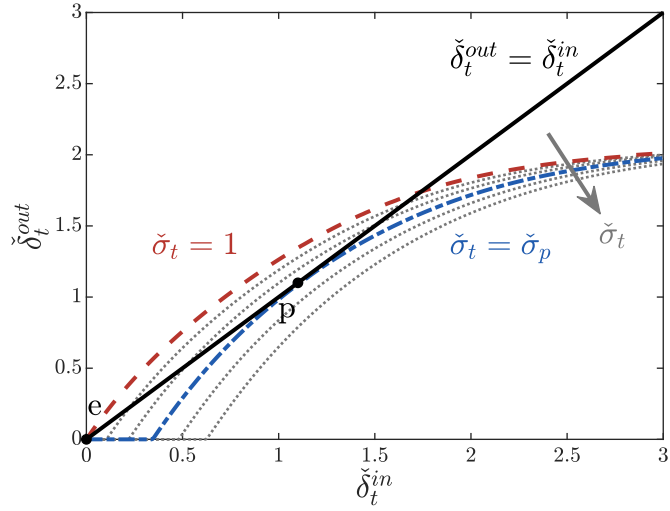}}
  \caption{Representation of the piecewise linear interpolation of pairs $(\check{\delta}_t^{in},\check{\delta}_t^{out})$ for increasing values of $\check{\sigma}_t$ starting from $\check{\sigma}_e=1$ (dotted lines). The plot is obtained for the \texttt{hE} class with profile shape $f(\tilde{x})=1-r\cdot(\tilde{x}-\check{\delta}_t^{in})$. Points \textit{e} and \textit{p} correspond to the solutions for pseudo-times $t=t_e$ and $t=t_p$, respectively.} 
  \label{fig:D_function2}
  \end{figure}

The boundary value problem can be reformulated as an initial value problem through the spatial coordinate transformation :
\begin{equation}
    \label{eq:coordinate_transformation}
    \tilde{x}=\check{x}+\check{\delta}_t^{in},
\end{equation}
where $\check{\delta}_t^{in}$ is a guess for the a priori unknown half-support 
width $\check{\delta}_t$. 
The damage criterion is rewritten in terms of $\tilde{x}$ as 
\begin{equation}
    \begin{aligned}
  \label{eq:dc_eta}
  \frac{1}{f_E(\tilde{x}-\check{\delta}_t^{in})}\frac{1}{(1-\tilde{\alpha}_t\left(\tilde{x})\right)^3}\,\check{\sigma}_t^2
  =
  f_w\left(\tilde{x}-\check{\delta}_t^{in}\right)\left(1-2\,\tilde{\alpha}_t''(\tilde{x})\right)+\\
  -2\,{f'_w}_{-}\left(\tilde{x}-\check{\delta}_t^{in}\right)\,\alpha_t'(\tilde{x})\hspace{3mm}\text{for }\hspace{3mm}\tilde{x}\geq0,
\end{aligned}
\end{equation}
where ${f_w'}_{-}(\check{x})$ is the left-hand derivative of ${f_w}(\check{x})$ and $\tilde{\alpha}(\tilde{x})$ is the function $\tilde{x}\mapsto\alpha_t(\tilde{x}-\check{\delta}_t^{in})$.
The initial conditions of the initial value problem stem from Eq.~\ref{eq:bcs_delta}:
\begin{equation}
    \label{eq:end_supp_eta}
    \tilde{\alpha}_t(0)=\tilde{\alpha}_t'(0)=0.
\end{equation}
 The initial value problem defined by Eqs.~\ref{eq:dc_eta}, \ref{eq:end_supp_eta} is solved via a Runge-Kutta scheme using the algorithm ODE 45 of \textsc{Matlab} \citep{shampine1997matlab} starting from the initial point $\tilde{x}=0$ (Figure~\ref{fig:D_function}). The integration is stopped at the target point $\tilde{x}=\check{\delta}_t^{out}\neq0$ such that 
\begin{equation}
    \label{eq:central_slope_eta}
    \tilde{\alpha}_t'(\check{\delta}_t^{out})=0\hspace{3mm}\text{for}\hspace{3mm}t>t_e.
\end{equation}
 For a given $\check{\sigma}_t$ with $t>t_e$, we assign to $\check{\delta}_t$ the values corresponding to the condition $\check{\delta}_t^{in}=\check{\delta}_t^{out}$ along the piecewise linear interpolation of the pairs $(\check{\delta}_t^{in},\check{\delta}_t^{out})$ obtained from Eqs.~\ref{eq:dc_eta}-\ref{eq:central_slope_eta}. When $t=t_e$, the bar is undamaged and $\check{\delta}_t=0$. For $1\leq\check{\sigma}_t\leq\check{\sigma}_p$, two values are assigned to $\check{\delta}_t$. The peak stress $\check{\sigma}_p$ is found as the stress for which these two values coincide (Figure \ref{fig:D_function2}).  Details about the numerical implementation are reported in~\ref{app:numericalDC}.
 
\par
\begin{figure}[H]
  \centering
     \makebox[\textwidth][c]{\includegraphics{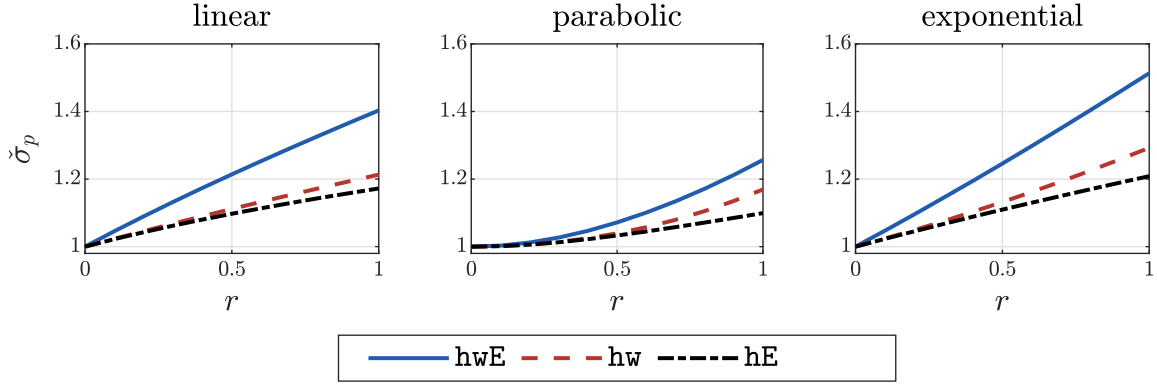}}
  \caption{Dimensionless peak stress vs.~characteristic ratio for different heterogeneity classes and profile shapes.}
  \label{fig:peak_lin_w1}
  \end{figure}

 Since the computation is based on Eqs.~\ref{eq:dc_eta}-\ref{eq:central_slope_eta}, the result, i.e.~the dimensionless peak stress $\check{\sigma}_p$, depends on the functions ${f_w}(\check{x})$ and ${f_E}(\check{x})$, i.e.~it depends on the heterogeneity class and profile shape and, for a given class and profile shape, it depends on the characteristic ratio $r$ only.
Results for all the considered heterogeneity classes and profile shapes are illustrated in Figure~\ref{fig:peak_lin_w1}.
 \begin{remark}
 \label{remark:hardening}
 For the heterogeneous bar, the peak stress $\sigma_p$ is larger than the elastic limit stress $\sigma_e$ (Figure~\ref{fig:peak_lin_w1}). This is due to the presence of a short hardening phase during the initial damage localization process.
 \end{remark}
\par
\begin{figure}[H]
    \hspace{1.5em}
    \begin{subfigure}[t]{0.6\textwidth}
        \centering
        \vspace{0.4em}
        \hspace{-3.5em}
        \raisebox{-\height}{\includegraphics[scale=1,trim=1mm 8mm 5mm 0.2mm]{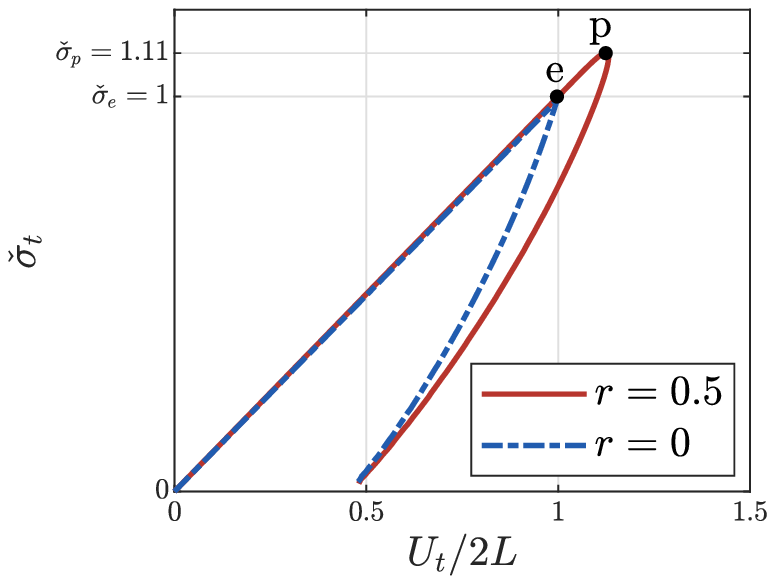}}
        \vspace{0.079\textwidth}
    \subcaption[]{} 
    \end{subfigure}%
    \hspace{-2em}
    \begin{subfigure}[t]{0.4\textwidth}
        \vspace{1em}
        \hspace{-0.1em}
        \raisebox{-\height}{\includegraphics{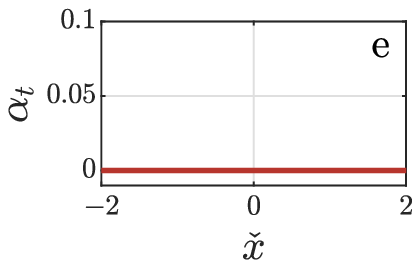}}
        \vspace{-0.03\textwidth}
        \subcaption[]{}
       \vspace{-0.07\textwidth}
       \hspace{-0.1em}
       \raisebox{-\height}{\includegraphics{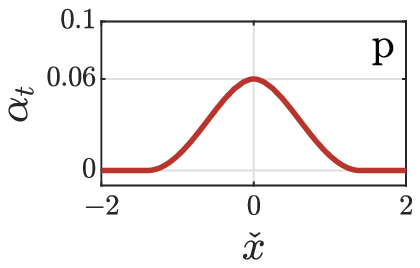}}
        \vspace{-0.03\textwidth}
        \subcaption[]{}
    \end{subfigure}
    \caption{Stress-displacement diagram for $\ell/2L=10^{-1}$, $\sqrt{\bar{w}_1/\bar{E}_0}=1$ for \texttt{hw} class with linear profile shape (a). For the case $r=0.5$, points \textit{e} and \textit{p} corresponding respectively to the pseudo-times $t=t_e$ and $t=t_p$ are indicated. Damage profile at $t=t_e$ (b). Damage profile at $t=t_p$ (c).}
  \label{fig:sd_lin_w1}
\end{figure}

 Computing the support extension for different values of $\check{\sigma}_t$ with $t\in [t_e,t_u)$ enables also the definition of the full stress-displacement curve during the damage localization phase. In this case, the numerical integration is performed via the stiff equation solver ODE 23s of \textsc{Matlab} \citep{shampine1997matlab}, see \ref{app:numericalDC} for the detailed algorithm.
To each $\check{\delta}_t$ we can associate a damage profile $\alpha_t(\tilde{x})$, its maximum $\alpha^*_t=\tilde{\alpha}_t(\check{\delta}_t)$, and the stress $\check{\sigma}_t$. Also, recalling Eq.~\ref{eq:sigma_epsilon}, by knowing the current dimensionless stress $\check{\sigma}_t$ and the damage profile $\alpha_t$, the dimensionless applied displacement $U_t/2L$ can be computed through the integral
\begin{equation}
    \label{eq:applied_displ}
    \frac{U_t}{2L}=\check{\sigma}_t\,\frac{\ell}{L}\,\sqrt{\frac{\bar{w}_1}{\bar{E}_0}}\,\int_{-L/\ell}^{0}\frac{1}{a(\alpha_t(\check{x}))\,f_E(\check{x})}\,d\check{x}.
\end{equation}
Sorting these quantities based on an ascending order of $\alpha^*_t$ yields the stress-displacement curve during the localization phase (Figure~\ref{fig:sd_lin_w1}).
 
 \begin{remark}
 \label{remark:revers}
In the spirit of previous studies on phase-field modeling of brittle fracture \textup{\citep{pham2011gradient, miehe2010thermodynamically}} and in compliance with the $\Gamma$-convergence arguments at the root of the approach, we performed here path-independent energy minimization, i.e. we do not account for the irreversibility condition. As noted in \textup{\citep{pham2013onset}}, the solution stemming from an incremental procedure and compatible with the irreversibility condition corresponds to the upper envelope of the set of localization profiles obtained for $t\in(t_e,t_u]$.
\end{remark}
 
\subsubsection{Fracture toughness} 
\label{subsct:FT}
The fracture toughness is defined as the dissipated energy at failure, and its determination requires the computation of the fully localized damage profile. At $t=t_u$ it is $\sigma_t=0$ and the damage criterion within the half-support width reads
\begin{equation}
    \label{eq:ode_lin}
    2\,\alpha_u''(\check{x})+2\,\frac{f'_w(\check{x})}{f_w(\check{x})}\alpha_u'(\check{x})=1\hspace{3mm}\text{in}\hspace{3mm} (-\check{\delta}_u,0).
\end{equation}
It is immediate to notice that, since function $f_E$ is only contained in the stress term which is now zero, the damage profile at $t=t_u$ for the heterogeneous bar only depends on function $f_w$. This implies that the fracture toughness for classes \texttt{hw} and  \texttt{hwE} depends on the profile shape of the heterogeneity and on the characteristic ratio $r$, whereas a heterogeneity of class \texttt{hE} leaves the fracture toughness  unchanged.

As follows, we exemplify the computations for the linear heterogeneity profile shape, whereas the analogous computations for the parabolic and exponential profile shapes follow similar lines and are reported in \ref{app:AT1_parabolic} and \ref{app:AT1_exponential}, respectively. For the linear profile shape, Eq.~\ref{eq:ode_lin} becomes

\begin{equation}
    \label{eq:ode_lin2}
    2\,\alpha_u''(\check{x})-2\left(\frac{r}{1-r\,\check{x}}\right)\alpha_u'(\check{x})=1\hspace{3mm}\text{in}\hspace{3mm} (-\check{\delta}_u,0).
\end{equation}
The analytical solution for $\alpha_u$ depends on the unknown coefficients $c_1$ and $c_2$:
\begin{equation}
    \label{eq:alpha_lin}
    \alpha_u(\check{x})= \frac{r\,\check{x}\left(-2+r\,\check{x}\right)+\left(-2+c_1\cdot 8\,r\right)\text{log}\left(1-r\,\check{x}\right)}{8r^2} + c_2\hspace{3mm}\text{in}\hspace{3mm} (-\check{\delta}_u,0).
\end{equation}
The two unknown coefficients can be obtained as functions of the unknown $\check{\delta}_u$ combining Eq.~\ref{eq:alpha_lin} with the boundary conditions Eq.~\ref{eq:bcs_delta} leading to
\begin{equation}
    \label{eq:c1_lin}
    c_1=-\frac{1}{4}\check{\delta}_u\left(2+r\,\check{\delta}_u\right),
\end{equation}
\begin{equation}
    \label{eq:c2_lin}
    c_2=\frac{-r\,\check{\delta}_u\,(2+\check{\delta}_u\, r)+2\,(1+r\,\check{\delta}_u)^2\,\text{log}(1+r\,\check{\delta}_u)}{8\,r^2}.
\end{equation}
By the following substitutions
\begin{equation}
    z=\frac{8\,r^2-1}{\text{exp}(1)}
    \hspace{3mm}\text{and}\hspace{3mm}
    \tau=\frac{8\,r^2-1}{(1+r\,\check{\delta}_u)^2}
\end{equation}
and using Eqs.~\ref{eq:alpha_lin}, \ref{eq:c1_lin}, \ref{eq:c2_lin}, we can rewrite the remaining boundary condition Eq.~\ref{eq:bcs2} as (\ref{app:nonlin_delta})
\begin{equation}
    \label{eq:delta_lambert}
    z=\tau\cdot \text{exp}(\tau).
\end{equation}
Eq.~\ref{eq:delta_lambert} has solution
\begin{equation}
    \tau=W_k(z),
\end{equation}
where $W_k$ is the \textit{Lambert function} whose definition is provided in \ref{app:Wfun}.
\par
\begin{figure}[H]
    \centering
    \hspace{1em}
    \begin{subfigure}{0.45\textwidth}
        \hspace{-1.1em}
    \includegraphics[scale=1,trim=1mm 8mm 5mm 0.2mm]{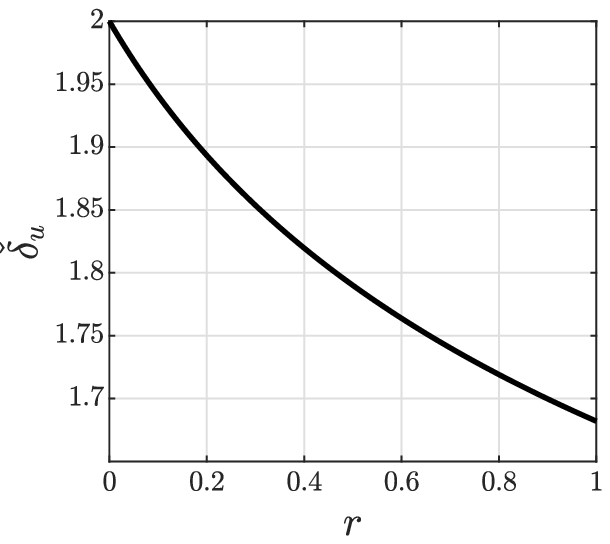}
    \vspace{1.3\baselineskip}
    \subcaption[]{}
    \label{subfig:delta}
    \end{subfigure}
    \hfill
    \begin{subfigure}{0.45\textwidth}
        \hspace{-1em}
    \includegraphics[scale=1,trim=1mm 8mm 5mm 0.2mm]{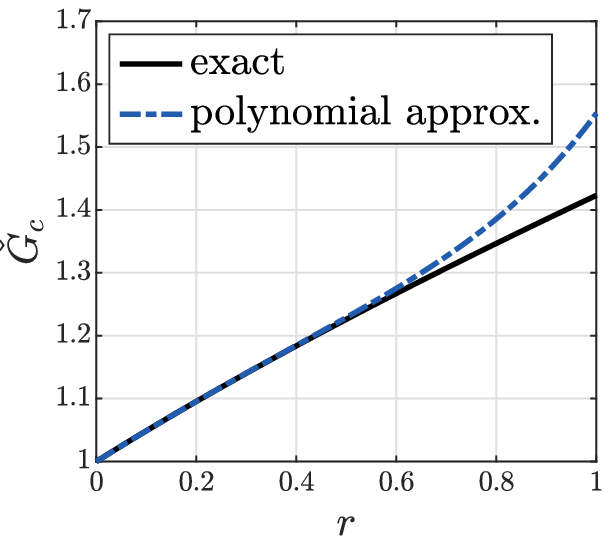}
    \vspace{1.3\baselineskip}
    \subcaption[]{}
    \label{subfig:Gc}
    \end{subfigure}
    \caption{Dimensionless half-support width at failure vs.~characteristic ratio for $f_w$ with linear shape (a). Dimensionless fracture toughness vs.~characteristic ratio for $f_w$ with linear shape (b).}
    \label{fig:delta_Gc}
\end{figure}

 Substituting backwards and prescribing $\check{\delta}_u\geq0$ we find
\begin{equation}
    \label{eq:delta_p}
    \check{\delta}_u=\frac{1}{r}\left(\,\text{exp}\left(\frac{1+W_0\left(\frac{8\,r^2-1}{\text{exp}(1)}\right)}{2}\right)-1\right)
\end{equation}
which gives the half-support width $\check{\delta}_u$ as a function of the characteristic ratio $r$ (Figure~\ref{subfig:delta}).
\par The dimensionless fracture toughness $\check{G}_c$, defined as
\begin{equation}
    \label{eq:dimless_G_c}
    \check{G}_c=\frac{G_c}{\bar{G}_c},
\end{equation}
where $\bar{G}_c=8/3\,\ell\,\bar{w}_1$ is the fracture toughness for the associated homogeneous material, can be obtained recalling Eqs.~\ref{G_c} and \ref{eq:energies_dam} as
\begin{equation}
    \label{eq:dimless_FT}
    \check{G}_c=\frac{3}{4}\int_{-\check{\delta}_u}^0 f_w(\check{x})\,\left(\alpha_u(\check{x})+\alpha_u'(\check{x})^2\right)d\check{x}.
\end{equation}
Combining Eq.~\ref{eq:alpha_lin} with Eqs.~\ref{eq:c1_lin} and \ref{eq:c2_lin} and using $f_w(\check{x})=h_{lin}(\check{x})$ in Eq.~\ref{eq:remapping}$_1$, $\check{G}_c$ can be expressed in terms of $\check{\delta}_u$ as
\begin{equation}
    \label{eq:Gc_delta}
    \begin{aligned}
    \check{G}_c= \frac{3}{256\, r^3}&\,(-3 - 4\,(1 + r\,\check{\delta}_u)^2\, (-1 + 2\,\text{log}(1 + r\,\check{\delta}_u)) + \\
    & + (1 + 
    r\,\check{\delta}_u)^4 (-1 + 4\, \text{log}(1 + r\,\check{\delta}_u))).
    \end{aligned}
\end{equation}
A further substitution of Eq.~\ref{eq:delta_p} in Eq.~\ref{eq:Gc_delta} leads to an analytical expression for $\check{G}_c$ that is plotted as solid line in Figure~\ref{subfig:Gc}.

\par
\begin{figure}[H]
    \centering
    \includegraphics{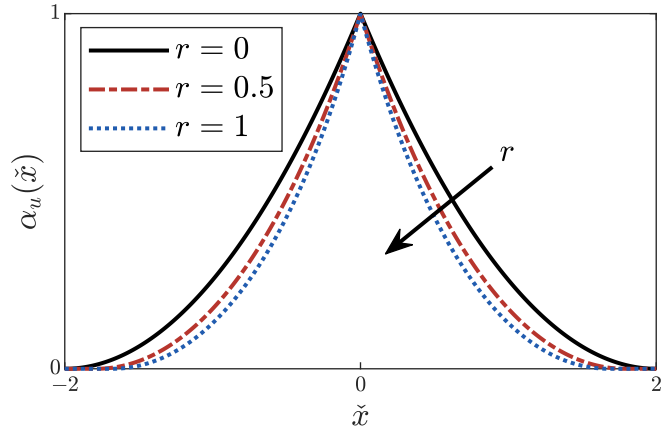}
    \caption{Damage profile at failure $\alpha_u$ for different values of the characteristic ratio for $f_w$ with linear shape.}
    \label{fig:alpha_r}
\end{figure}

 The expression for $\check{G}_c$ can be simplified using the polynomial approximation of the Lambert function proposed by Veberi{\v c} \citep{veberivc2012lambert} along with a Taylor expansion about $r=0$. Although the approximation order can be freely selected (\ref{app:lambert}), here the Veberi{\v c} approximation is truncated at order 6 and the Taylor expansion at order 5 giving 
\begin{equation}
    \label{lin_veberic}
    \check{G}_c\approx1+\frac{1}{2}r-\frac{2}{15}r^2+\frac{16}{135}r^3-\frac{46}{315}r^4+\frac{608}{2835}r^5+o(r^5).
\end{equation}
The exact dimensionless fracture toughness (Eq.~\ref{eq:Gc_delta}) and its polynomial approximation (Eq.~\ref{lin_veberic}) are compared in Figure~\ref{subfig:Gc}. Finally, combining Eqs.~\ref{eq:alpha_lin},~\ref{eq:c1_lin},~\ref{eq:c2_lin},~\ref{eq:delta_p} we can plot the damage profile at failure, $\alpha_u$, for different values of the characteristic ratio $r$ (Figure~\ref{fig:alpha_r}).

\begin{figure}[H]
  \centering
     \makebox[\textwidth][c]{\includegraphics{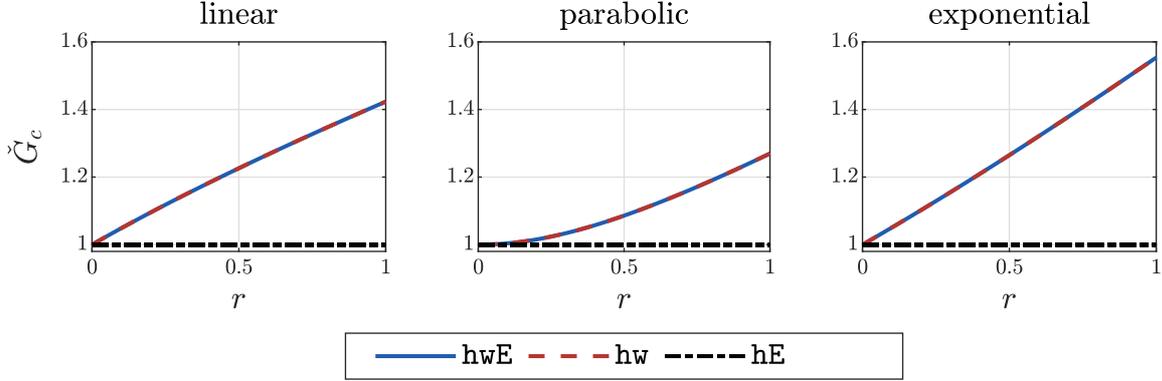}}
  \caption{Dimensionless fracture toughness vs.~characteristic ratio for different heterogeneity classes and profile shapes.}
  \label{fig:results_FT}
  \end{figure}

The results on the dimensionless fracture toughness for the three different profile shapes are summarized in Figure \ref{fig:results_FT}.

 \begin{remark}
 For the heterogeneous bar, the damage profile at failure is narrower and the fracture toughness $G_c$ is larger than for the bar made of the associated homogeneous material (Figures~\ref{fig:alpha_r},  \ref{fig:results_FT}).
 \end{remark}

\subsection{Extension to heterogeneous bars in three dimensions}
\label{subsct:3D}
In this subsection, we extend the previous one-dimensional analysis to the case of bars with properties varying along the longitudinal axis embedded in the three-dimensional space. The main aim is to investigate the effects of this more realistic setting on the peak stress and fracture toughness and highlight the differences with the one-dimensional case (Sections~\ref{subsct:peak} and  \ref{subsct:FT}). 

\subsubsection{Problem setting}
\par
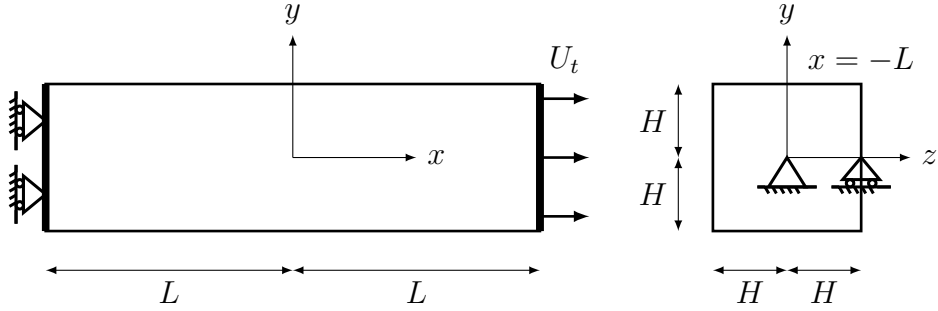
\begin{figure}[H]
    \centering
    \begin{tikzpicture}[scale = 0.65]
	\draw[line width = 1] (0, 0) rectangle (10, 3);
    \draw[-latex] (5, 1.5) -- +(2.5, 0) node[right, scale = 1]{$x$};
	\draw[-latex] (5, 1.5) -- +(0, 2.5) node[above, scale = 1]{$y$};
	\foreach \nn in {0.3, 1.5, ..., 2.7} {
    \draw[-latex] [ line width = 1] (10, \nn) -- (11, \nn);}
    \node[scale = 1.] (nodeU) at (10.5, 3.5) {$U_t$};
    \draw [latex-latex] (0,-0.8) -- (5, -0.8) node[midway, below, scale = 1.]{$L$};
    \draw [latex-latex] (5,-0.8) -- (10, -0.8) node[midway, below, scale = 1.]{$L$};
    \draw[line width=1 mm] (10,0) -- (10,3);
	\foreach \nn in {0.75, 2.25} {
	\RollerSupport[270]{0,\nn}{1.5};}
	\draw[line width=1 mm] (0,0) -- (0,3);
	
	\draw[line width = 1] (13.5, 0) rectangle (16.5, 3);
    \draw[-latex] (15, 1.5) -- +(2.5, 0) node[right, scale = 1]{$z$};
	\draw[-latex] (15, 1.5) -- +(0, 2.5) node[above, scale = 1]{$y$};
	\node[scale = 1.] (nodeU) at (16.5, 3.5) {$x=-L$};
	\RollerSupport[0]{16.5,1.5}{1.5};
	\HingeSupport[0]{15,1.5}{1.5};
    \draw [latex-latex] (13.5,-0.8) -- (15, -0.8) node[midway, below, scale = 1.]{$H$};
    \draw [latex-latex] (15,-0.8) -- (16.5, -0.8) node[midway, below, scale = 1.]{$H$};
    \draw [latex-latex] (12.8,0) -- (12.8, 1.5) node[midway, left, scale = 1.]{$H$};
    \draw [latex-latex] (12.8,1.5) -- (12.8, 3) node[midway, left, scale = 1.]{$H$};
    \draw[line width=1 mm] (10,0) -- (10,3);
	
    \end{tikzpicture}
    \caption{\label{setting3D} {Geometry and boundary conditions in three dimensions.}}
\end{figure}

 Figure~\ref{setting3D} illustrates the geometry and the boundary conditions of the problem. The bar has length $2L$ and square cross-section with dimensions $2H\times2H$, and a monotonically increasing displacement $U_t$ is prescribed at $x=L$. Homogeneous Neumann boundary conditions are enforced to the damage variable on the whole boundary.
\par The three-dimensional version of the total energy functional in Eq.~\ref{eq:tot_en_fun} reads
\begin{equation}
     \mathcal{E}(\boldsymbol{u},\alpha)=\int_{\Omega}\left(\frac{1}{2}\,a(\alpha)\,\mathbb{C}_0 \boldsymbol{\varepsilon}(\boldsymbol{u})\cdot \boldsymbol{\varepsilon}(\boldsymbol{u})+w_1\left(w(\alpha)+\ell^2 \,\vert\nabla\alpha \vert^2 \right)\right)\,d\boldsymbol{x},
\end{equation}
where $\boldsymbol{u}$ is the displacement vector, $\mathbb{C}_0$ is the undamaged 4th-order elasticity tensor for linear elastic isotropic  materials, $\boldsymbol{\varepsilon}(\boldsymbol{u})=\frac{1}{2}(\nabla\boldsymbol{u}+\nabla^T\boldsymbol{u})$ is the 2nd-order infinitesimal strain tensor, $\Omega$ is the spatial domain and $\boldsymbol{x}=\{x,y,z\}$ is the spatial coordinate vector with $x$ corresponding to the longitudinal axis of the bar. 
\par The distribution of the elastic and fracture properties along the major axis is defined again through the functions $f_E$ and $f_w$ as
\begin{equation}
    \label{eq:het_3d}
    \mathbb{C}_0(\boldsymbol{x})=\bar{\mathbb{C}}_0\cdot f_E(x)\hspace{3mm}\text{and}\hspace{3mm}w_1(\boldsymbol{x})=\bar{w}_1\cdot f_w(x),
\end{equation}
where $\bar{\mathbb{C}}_0$ and $\bar{w}_1$ are independent of $\boldsymbol{x}$.
In \ref{app:mat_par} it is shown that, for isotropic material properties, Eq.~\ref{eq:het_3d} implies a heterogeneous distribution of the undamaged elastic modulus, namely $E_0(\boldsymbol{x})=\bar{E}_0\cdot f_E(x)$, and a homogeneous Poisson's ratio $ \nu(\boldsymbol{x})=\bar \nu$.

\par The main hypotheses adopted in the one-dimensional case are preserved. In particular, we perform a path-independent energy minimization (Remark~\ref{remark:revers}), we introduce the dimensionless spatial coordinate vector $\check{\boldsymbol{x}}=\boldsymbol{x}/\ell$ and the dimensionless stress tensor  $\check{\boldsymbol{\sigma}}_t=\boldsymbol{\sigma}_t/\sqrt{\bar{w}_1\,\bar{E}_0}$. We limit the analysis to a linear type of heterogeneity with slope $r$ (Table \ref{tab:shapes}). 

Equilibrium prescribes {$\div(\check{\boldsymbol{\sigma}}_t)=\boldsymbol{0}$}, while the damage criterion is written as
\begin{equation}
    \label{eq:3d_problem}
    \frac{1}{f_E}\frac{1}{(1-\alpha)^3}\,\check{\mathbb{S}}_0\check{\boldsymbol{\sigma}}_t\cdot \check{\boldsymbol{\sigma}}_t
  =
  f_w\cdot \left(1-2\,\Delta\alpha_t\right)-2\,\frac{\partial f_w}{\partial\check{x}}\frac{\partial{\alpha}_t}{\partial\check{x}}\hspace{3mm} \text{in}\hspace{3mm}\check{\mathcal{S}}_{t},
\end{equation}
where $\check{\mathcal{S}}_{t}$ is the support of the damage variable. $\check{\mathbb{S}}_0$ is the dimensionless 4th-order undamaged compliance tensor, which reads
\begin{equation}
    \label{}
    \check{\mathbb{S}}_0=\bar{E}_0\,\bar{\mathbb{C}}_0^{-1}=(1+\nu)\,\mathbb{I}^s-\nu\,\boldsymbol{I}\otimes\boldsymbol{I}
\end{equation}
and depends on the Poisson's ratio $\nu$ only. 
Note that Eq.~\ref{eq:3d_problem} depends exclusively on the parameters $r$ and $\nu$ and on the class of heterogeneity (\texttt{hw}, \texttt{hE}, \texttt{hwE}). 
Thus, in a three-dimensional setting the damage criterion depends on the Poisson's ratio and the criterion defining the elastic limit $t_e$ becomes $\max_{\check{\boldsymbol{x}}}(\check{\mathbb{S}}_0 \check{\boldsymbol{\sigma}}_e(\check{\boldsymbol{x}})\cdot \check{\boldsymbol{\sigma}}_e(\check{\boldsymbol{x}}))
  =1$.

\subsubsection{Peak average stress}
\par In three dimensions the stress can vary within the cross-section. However, equilibrium imposes a constant axial force, hence a constant cross-sectional $average$ normal stress $\langle\sigma\rangle_t$, which is defined as
\begin{equation}
    \langle\sigma\rangle_t=\frac{1}{4\,H^2}\int_{-H}^{H}\int_{-H}^{H}\sigma_{{xx}_t}\,dy\,dz,
\end{equation}
where $\sigma_{{xx}_t}$ is the $xx$- component of the Cauchy stress tensor $\boldsymbol{\sigma}_t=a(\alpha_t)\,\mathbb{C}_0 \boldsymbol{\varepsilon}(\boldsymbol{u}_t)$. Therefore, in the three-dimensional context the concept of peak stress used in Section~\ref{subsct:peak} is replaced by the average peak stress $\langle\sigma\rangle_p=\text{sup}_t\langle\sigma\rangle_t$. 
\par Thus, let us study the peak average dimensionless stress $\langle \check{\sigma}\rangle_p=\langle \sigma\rangle_p/\sqrt{\bar{w}_1\bar{E}_0}$. We perform a series of numerical experiments for different value of $r\in[0,1]$ and four values of the Poisson's ratio $\nu=\{0,0.15,0.3,0.45\}$. We assume to have a long bar with $L/\ell=15$,  aspect ratio $L/H=6.25$, $\ell=0.1$, $\bar{E}_0=1$ and $\bar{w}_1=1$. The finite element analyses are carried out with our code \texttt{GRIPHFiTH} \citep{griphfit2022} using a structured mesh with 8-node brick elements. Along $y$ and $z$, the mesh size is $0.02$. The mesh is refined in the central zone along the $x$-axis for a length of $4\,\ell$, i.e. where the damage is expected to localize. In this central region, the element size is $0.01$, while outside it is $1.3/7$. The load $U_t$ is applied in steps with increments of $\Delta U_t=5\cdot10^{-4}$. 
The non-negativity of the damage variable is enforced through the penalty method proposed by Gerasimov et al.~\citep{gerasimov2019penalization} and the problem is solved by means of an alternate minimization scheme \citep{bourdin2007numerical} which is stopped when the norm of the residuals becomes lower than a predefined tolerance of 10$^{-4}$.

\begin{figure}[H]
  \centering
     \makebox[\textwidth][c]{\includegraphics{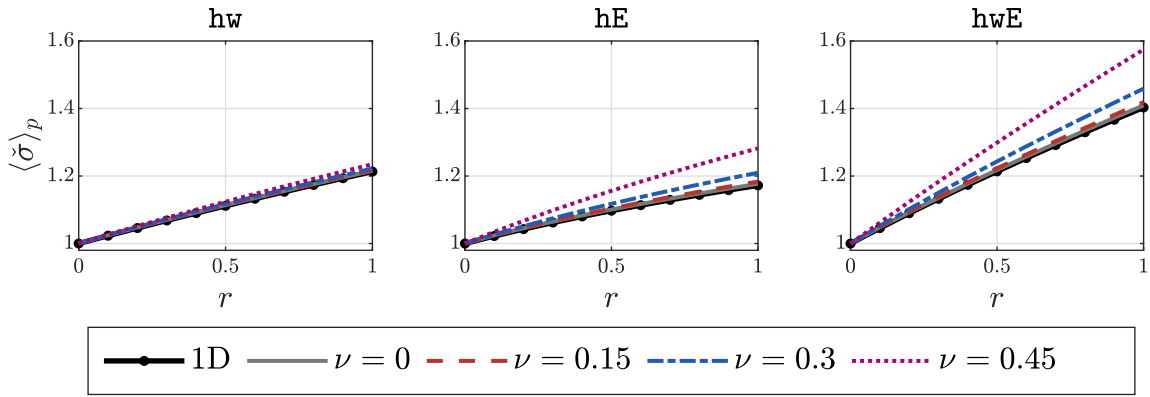}}
  \caption{Peak average dimensionless  stress vs.~characteristic ratio for different values of the Poisson's ratio $\nu$ and different classes of heterogeneity with linear profile shape.}
  \label{fig:sp_nu_dep}
  \end{figure}

The results in terms of $\langle\check{\sigma}\rangle_p$ for the three classes of heterogeneity are summarized in Figure~\ref{fig:sp_nu_dep} and are qualitatively very similar to those obtained in the one-dimensional setting, showing an increase of the peak stress with the characteristic ratio $r$. 

\par Quantitatively we observe that the values obtained for $\nu=0$ overlap with those of the one-dimensional bar (as expected), while increasing the Poisson's ratio leads to an increase of the  peak average stress $\langle \check{\sigma}\rangle_p$. For the classes \texttt{hE} and \texttt{hwE}, this is due to a varying lateral contraction introduced by the Poisson's effect whose magnitude is modulated by the elastic modulus along the bar. As a consequence, a necking deformation is introduced already in the elastic regime whose magnitude varies along the $x$-axis and has its maximum in correspondence of the minimum value of $\bar E_0(\boldsymbol{x})$. It is worth noting that this necking deformation is not present in homogeneous bars, at least before the onset of the failure after reaching the peak stress $\sigma_p=\sigma_e$. The change in lateral contraction introduces parasitic shear stresses that break up the uniaxial stress state, leading to a variation of the term $\check{\mathbb{S}}_0\check{\boldsymbol{\sigma}}_t(\check{\boldsymbol{x}})\cdot \check{\boldsymbol{\sigma}}_t(\check{\boldsymbol{x}})$ within the cross-section and, thus, to a point-wise different local elastic limit. This effect is more pronounced for larger values of the Poisson's ratio and is further exacerbated by the progressive evolution of the damage during the hardening regime (see Section~\ref{subsct:peak}), which makes the heterogeneity in the elastic parameters more pronounced. As a result of this mechanism the damage does not evolve homogeneously within the cross-section, and this ultimately leads to an increase of the  peak average stress with respect to the one-dimensional case. 
\par
\begin{figure}
\centering
\begin{subfigure}{0.75\textwidth}
    \includegraphics[scale=1]{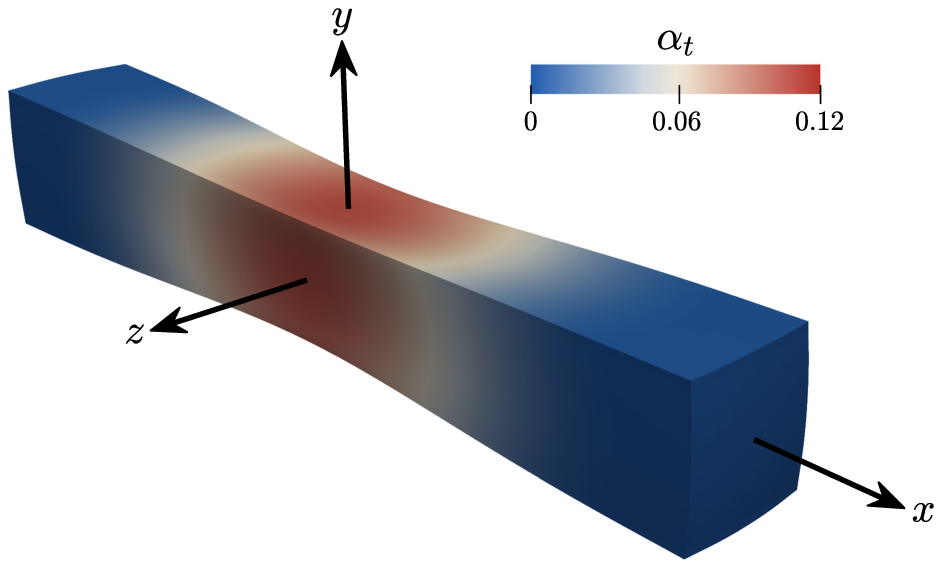}
    \subcaption[]{}
    \label{subfig:necking}
\end{subfigure}
\begin{subfigure}{0.21\textwidth}
    \includegraphics[scale=1]{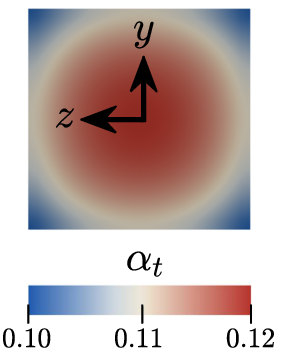}
    \subcaption[]{}
    \label{subfig:cross_section}
\end{subfigure}
\caption{Necking of a bar with $r=0.5$, $\nu=0.3$ and class \texttt{hw} at the last time step before failure. The necking is due to the localization of the damage $\alpha_t$. In order to emphasize the necking effect, we only visualize the central portion of the bar from $x=-2\,\ell$ to $x=2\,\ell$ (a). Inhomogeneous damage field over the cross-section at $x=0$ (b).}
\label{fig:hard_3Ds}
\end{figure}

 Although to a limited extent, this effect is also observed in absence of heterogeneity in the elastic properties, as demonstrated by the results for class \texttt{hw} (Figures~\ref{fig:sp_nu_dep} and \ref{subfig:necking}). In this case, the elastic limit is the same for all the points within the cross-section where localization is expected, however the uniaxial stress state is altered by the evolution of the damage during the hardening regime, which promotes a spatial variation of the elastic parameters and an inhomogeneous distribution of the damage variable within the cross-section (Figure~\ref{subfig:cross_section}). 

\subsubsection{Fracture toughness}
\par At failure, the left-hand side of Eq.~\ref{eq:3d_problem} vanishes since $\boldsymbol{\sigma}_u=\boldsymbol{0}$, thus, the dependence of the problem on the Poisson's ratio disappears and the three-dimensional problem simplifies in a scalar one-dimensional-like problem.  As a consequence the dimensionless fracture toughness depends only on $r$ and on the function $f_w$ and the same observations reported in Section \ref{subsct:FT} apply also here.

\par
\begin{figure}[H]
  \centering
     {\includegraphics[scale=1]{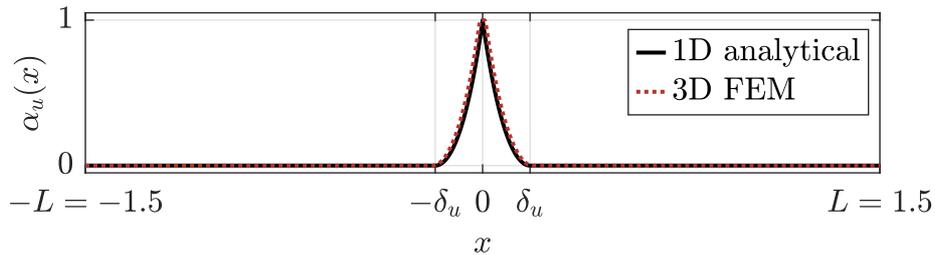}}
  \caption{Damage at failure along $x$ obtained for the three-dimensional bar  with $r=0.5$, $\nu=0.3$ and class \texttt{hw} using the finite element method (FEM). The numerical damage profile overlaps with the analytical profile derived for the one-dimensional bar.}
  \label{fig:alpha_f_3D}
  \end{figure}

 This result is also verified numerically. After the peak load, the system enters a softening branch characterized by a snap-back where the damage evolves abruptly until reaching the complete failure of the bar, i.e. $\alpha_u=1$ at $x=0$. At this point the damage is homogeneous within the cross-section and its profile obtained numerically coincides with the analytical result derived in the one-dimensional case (Figure~\ref{fig:alpha_f_3D}).

\subsection{Discussion}
\label{subsct:discussion}
In this subsection we discuss some implications of the obtained results.
\subsubsection{Non-locality}
\label{sct:non_loc_disc}
In the present section we have investigated how heterogeneity in the elastic and fracture material properties affects the observed behavior of a bar. We have assumed the material properties to be minimum at the midpoint cross-section of the bar (we have taken these as reference material properties). As a result, the midpoint cross-section is the location where the elastic limit is reached first and around which damage localization starts and develops, finally leading to failure of the bar.

We have concluded that, for a bar with a given heterogeneity class and profile shape, the peak stress $\sigma_p$ and the fracture toughness $G_c$ can be expressed as 
\begin{equation}
    \label{eq:r_only}
    \sigma_p=\bar{\sigma}_p\cdot p(r), \hspace{10mm} G_c=\bar{G}_c\cdot g(r), 
\end{equation}
where $\bar{\sigma}_p$ and $\bar{G}_c$ denote the peak stress and the fracture toughness of the bar made of the homogeneous reference material, and $p(r):r\mapsto\check{\sigma}_p$ and $g(r):r\mapsto\check{G}_c$ have been determined to be always larger than $1$ and increasing with $r$. Thus the peak stress and fracture toughness of the heterogeneous bar are both larger than those of the homogeneous reference bar. This increase is a non-local effect resulting from the elastic modulus and/or the specific fracture energy being larger than those of the reference homogeneous material in the neighborhood of the midpoint cross-section, thus it is a consequence of the choice made for the reference homogeneous material. The non-locality is naturally induced by the phase-field model through its intrinsic length scale, so that the macroscopic behavior of the bar does not simply result from the local properties at the midpoint cross-section but involves its neighborhood. Accordingly, the increase in peak stress and fracture toughness is a function of the ratio $r$ between the internal length of the phase-field model and the length characterizing the speed of variation of the material properties. The non-local effect vanishes, i.e.~$p(r)$ and $g(r)$ approach $1$, when the limit case $r\rightarrow 0$ is approached, \ie~for the sharp-crack model (or, trivially, for homogeneous material properties).

\subsubsection{Model calibration}
For calibration of the phase-field model, different options are possible depending on which properties can be realistically assumed to be known. Let us assume that the shape of the heterogeneity in the material properties  is known upfront, e.g. through computed tomography by correlation with the density profile. This is a common practice in many fields, \eg~bone biomechanics \citep{currey1988effect, katz2019scanner} (however correlation is typically assumed between the density and the value of the elastic modulus, whereas the analogous correlation with the specific fracture energy is less investigated). Under this assumption, $\ell_f$ is also known. To fix ideas, let us further assume that the heterogeneity shape corresponds to one of the profile shapes we considered in this study. Quantities which can be realistically measured on the bar geometry are the initial stiffness $k_0$, the elastic limit stress $\sigma_e$ and the peak stress $\sigma_p$. The initial stiffness can be expressed as
\begin{equation}
    \label{eq:compliance}
    k_0=\frac{1}{\int_{-L}^{L}E_0(x)^{-1}dx}
\end{equation}
from which the value of the elastic modulus at the midpoint cross-section, $\bar{E}_0$, can be deduced. From Eq.~\ref{eq:sigma_e}, $\bar{w}_1$ can be computed from the measurement of $\sigma_e$. Finally, from the measurement of $\sigma_p$, the intrinsic length $\ell$ of the phase-field model can be calibrated using Eq.~\ref{eq:r_only}$_1$ and recalling that $\bar{\sigma}_p=\bar{\sigma}_e=\sigma_e$. 



\section{Sharp-crack model vs.~phase-field model}
\label{subsct:SvP}
Let us now explore further the consequences of the non-local nature of the phase-field model, as opposed to the locality of the sharp-crack model, in bars made of heterogeneous materials.

\subsection{Sharp-crack model}
The sharp-crack model, put forth by Francfort and Marigo \citep{francfort1998revisiting} as a variational reformulation of Griffith's brittle fracture criterion, relies on the \emph{global} minimization of a total energy functional. 
In the one-dimensional case, this functional can be expressed as \citep{gerasimov2019penalization}
\begin{equation}
    \mathcal{E}_{\texttt{sc}}(u,\Gamma_c):=\int_{\Omega\setminus\Gamma_c}\frac{1}{2}\,E_0\,u'^2\,dx+\int_{\Gamma_c}\bar{G}_c\, f_w(x)\,\mathbb{H}^0(dx),
\end{equation}
where $\Omega$ is the problem domain, $\Gamma_c$ is the \emph{crack set}, i.e.~the set including the cracked points of the bar, and $\mathbb{H}^0(\Gamma_c)$ is its Hausdorff measure which, in the one-dimensional case, returns  the number of points belonging to $\Gamma_c$. The product $\bar{G}_c\, f_w(x)$ can be regarded as the \emph{specific fracture energy} for the sharp-crack model. Accordingly, also in this case $f_w(x)$ plays the role of spatial variation profile of the fracture property and $\bar{G}_c$ represents the minimum value of the specific fracture energy.
While the phase-field model is based on \textit{local} energy minimization (see the local stability condition in Problem~\ref{problem}), the sharp-crack model needs the minimization of the total energy functional to be \textit{global}, otherwise crack nucleation would not be predicted since the undamaged elastic solution is always locally stable. Within this global minimization framework, an initially uncracked bar under tension remains sound until the creation of a crack becomes energetically more convenient than the storage of additional elastic energy. At critical conditions, the stored elastic energy is released completely and the fracture energy takes the value corresponding to a single crack formed in correspondence of the minimum value of  $f_w(x)$ \citep{francfort1999cracks}. As claimed in \citep{francfort1998revisiting}, this result marks a contrast with classical fracture mechanics based on Griffith's criterion, as it amends its inability to predict crack initiation. 

\subsection{Multiple minima for $f_w$}
\par An interesting benchmark showcasing the differences between sharp-crack and phase-field models is the case of bars where multiple points compete as initiation sites, namely when \emph{multiple minima} for $f_w$ are present. In the following, we present two different examples, one with two equal minima and one with two different minima. 

\subsubsection{Two equal minima for $f_w$}
\label{subsubsct:2gm}
\par
\begin{figure}[H]
\centering
\begin{subfigure}[b]{0.9\textwidth}
    \hspace{-0.62em}
   \includegraphics[scale=1]{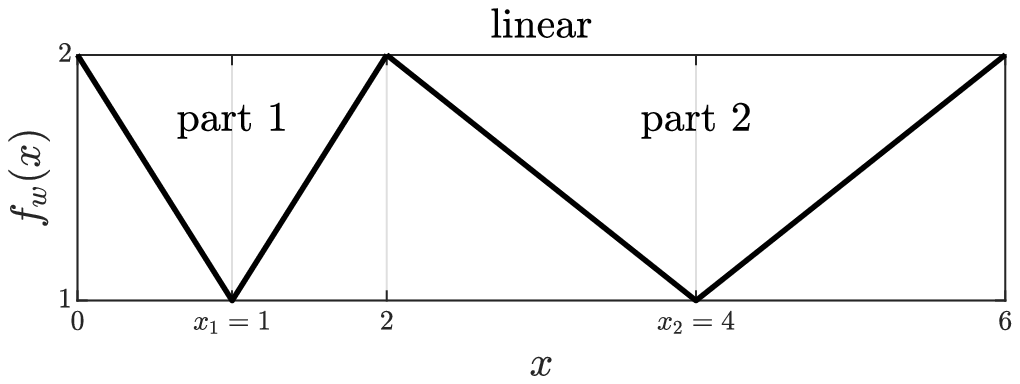}
   \caption{}
   \label{subfig:pline} 
\end{subfigure}

\begin{subfigure}[b]{0.9\textwidth}
    \hspace{-0.62em}
   \includegraphics[scale=1]{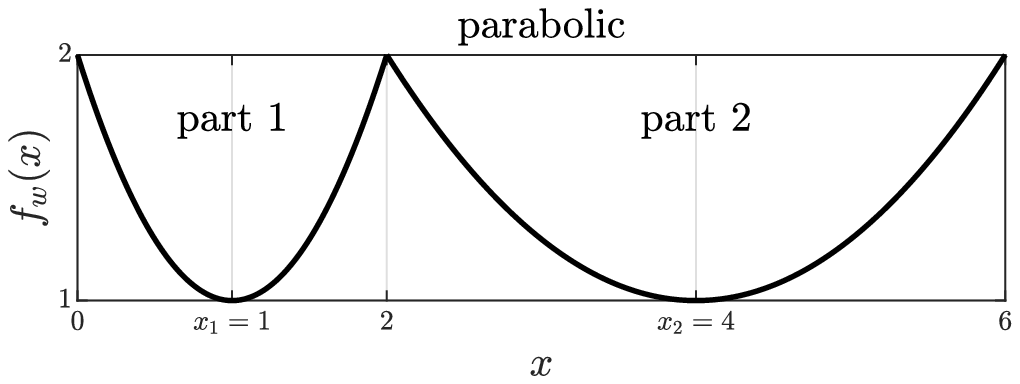}
   \caption{}
   \label{subfig:pparabola}
\end{subfigure}

\caption{Linear (a) and parabolic (b) examples of heterogeneity profile with two equal minima at $x_1$ and $x_2$. The profiles can be split into a first part symmetric about $x_1$ and a second part symmetric about $x_2$.}
\label{fig:asymmetric}
\end{figure}

 We consider heterogeneity in specific fracture energy with a profile $f_w(x)$ possessing two equal minima in $x_1$ and $x_2$. We further assume that the profile can be split into two parts, one symmetric about $x_1$ and the other symmetric about $x_2$, and we consider linear and parabolic profiles as in Figure~\ref{fig:asymmetric}. We also introduce the function $f_{w,1}(x-x_1)$  describing the right-half of the first part of the profile and $f_{w,2}(x-x_2)$ describing the right-half of the second part of the profile. For the examples in Figure~\ref{fig:asymmetric}, we have:
\begin{align}
\label{eq:examples}
  \text{linear }
  \begin{cases}
    f_{w,1}(x)=1+x \\
    f_{w,2}(x)=1+\frac{1}{2}\,x
\end{cases},\,
  \text{parabolic }
  \begin{cases}
    f_{w,1}(x)=1+x^2 \\
    f_{w,2}(x)=1+\frac{1}{4}\,x^2
\end{cases}.
\end{align}
\par Let us first study this problem using the sharp-crack model. Gerasimov et al. \citep{gerasimov2020stochastic} showed that, for this profile, the problem with the sharp-crack model is ill-posed due to the competition between the two possible crack locations $x_1$ and $x_2$. They proposed a stochastic relaxation to transform the ill-posed deterministic problem into a well-posed stochastic problem formulated in terms of fracture probability. The stochastic solution was found by introducing a random perturbation to the specific fracture energy profile, in form of a white noise with magnitude controlled by the small parameter $\eta>0$, and then letting $\eta$ approach $0$. Denoting with $P_i$ the probability that the crack forms at $x_i$ ($i=1,2$), it was concluded that, for the linear and parabolic examples in Figure~\ref{fig:asymmetric}, $P_1=1/3$ and $P_2=2/3$ \citep{gerasimov2020stochastic}.

\par In \citep{gerasimov2020stochastic}, the mentioned probabilities are calculated numerically but it is possible to obtain the same results in closed form. 
For a given parameter $\eta$ controlling the magnitude of the noise, the number of favorable cases for fracture at $x_i$ is proportional to $f_{w,i}^{-1}(1+\eta)$, whereas the number of possible cases is proportional to $\sum_{j=1}^{2}f_{w,j}^{-1}(1+\eta)$. Therefore,
\begin{equation}
    \label{eq:prob_approach}
    P_i=\lim_{\eta\rightarrow 0}\frac{f_{w,i}^{-1}(1+\eta)}{\sum_{j=1}^{2} f_{w,j}^{-1}(1+\eta)},
\end{equation}
Eq.~\ref{eq:prob_approach} can be easily extended to the case where the profile is not symmetric and to an arbitrarily large number of minima. 
Both for linear and parabolic examples, Eq.~\ref{eq:prob_approach} together with Eq.~\ref{eq:examples} yields $P_1=1/3$ and $P_2=2/3$ which coincide with the results in \citep{gerasimov2020stochastic}.

\par Let us now solve the same problem using the phase-field modeling approach. Exploiting the results in the present study, it is straightforward to demonstrate that the problem becomes well-posed. For the profiles in Figure~\ref{fig:asymmetric}, for a given value of $\ell$, the characteristic ratio $r_1$ about $x_1$ is larger than the characteristic ratio $r_2$ about $x_2$. Therefore, according to the result in Figure~\ref{fig:peak_lin_w1}, fracture at $x_1$ requires a larger stress than fracture at $x_2$, hence, fracture can only occur at $x_2$. 
\par
\begin{figure}[H]
  \hspace{-0.95em}
  \makebox[\textwidth][c]{\includegraphics{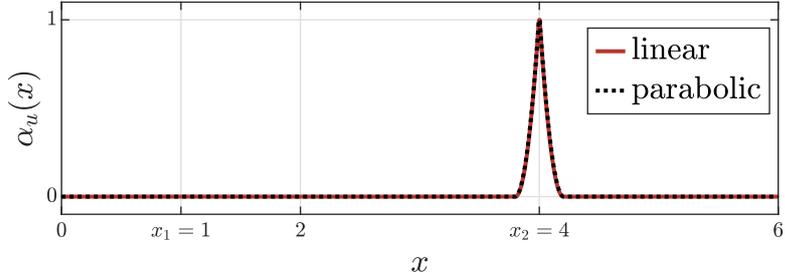}}
    \caption{Phase-field damage profile for the linear and parabolic profile examples (with $\ell=0.1$). In both cases, a crack at $x_2$ is predicted.}
  \label{fig:alpha_e1_e2}
  \end{figure}

 This is confirmed by a one-dimensional finite element analysis performed assuming $\ell=0.1$, $\bar{E}_0=1$, $\bar{w}_1=1$ and no heterogeneity in the elastic modulus. We use a uniform mesh with linear elements of size $\ell/30$. At every time step, the prescribed end displacement is increased by a constant increment $\Delta U_t=0.1$. Irreversibility is enforced through the penalty method \citep{gerasimov2019penalization} and the same alternate minimization scheme mentioned in the three-dimensional analysis is adopted. In Figure~\ref{fig:alpha_e1_e2}, the results at failure are shown. As expected, in both cases the crack forms at point $x_2$. 

Thus, in this example the non-local regularization inherent to the phase-field model amends the ill-posedness of the problem. Since the non-local model "sees" not only the minimum in the specific fracture energy but also its neighborhood, equal minima surrounded by different neighborhoods (which reflects in unequal values of $r$) lead to different peak stresses and the phase-field regularized problem becomes well-posed. 

\subsubsection{Two different minima for $f_w$}

 \par
\begin{figure}[H]
  \centering
    \hspace{-0.62em}
     \makebox[\textwidth][c]{\includegraphics{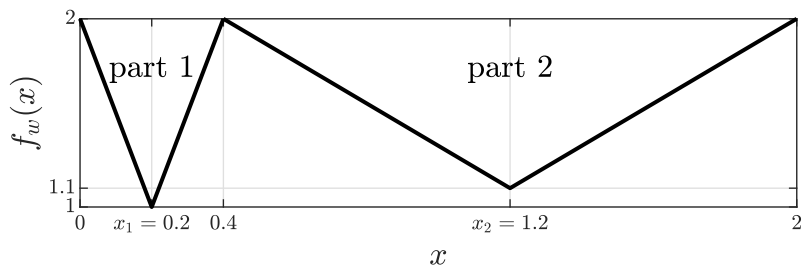}}
  \caption{Example of heterogeneity profile with two minima at $x_1$ and $x_2$ with two different values $f_w(x_1)=1$ and $f_w(x_2)=1.1$. The profile can be split into a first part symmetric about $x_1$ and a second part symmetric about $x_2$.}
  \label{fig:diff_loc_min}
  \end{figure}

 Let us now study an example with two different minima at $x_1$ and $x_2$ (Figure~\ref{fig:diff_loc_min}).
This time, according to the sharp-crack model the problem is well-posed. Indeed, the crack is predicted to initiate at $x_1$, at which $f_w(x)$ possesses its global minimum \citep{francfort1999cracks}.
\par
\begin{figure}[H]
  \centering
  \hspace{-0.62em}
     \makebox[\textwidth][c]{\includegraphics{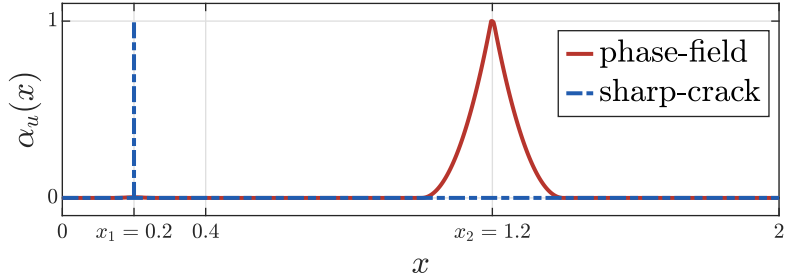}}
  \caption{Results for the example with two different minima: the sharp-crack model predicts a crack at $x_1$, the phase-field model (with $\ell=0.1$) predicts a crack at $x_2$.}
  \label{fig:alpha_prof_diff}
  \end{figure}

 In order to solve this problem with the phase-field approach, we assume that the material is heterogeneous in $w_1$ only and that $\ell=0.1$. Accordingly, $r_1=0.5$ and $r_2=0.1125$. We can then estimate the peak stress for a fracture at $x_i$ with the relationship $\sigma_{p_i}=\check{\sigma}_{p_i}\sqrt{\bar{E}_0\,\bar{w}_1\,f_w(x_i)}$, where $\check{\sigma}_{p_i}$ is the dimensionless peak stress for linear shape and class \texttt{hw} extracted from Figure~\ref{fig:peak_lin_w1} at $r=r_i$. We find that $\sigma_{p_1}\simeq1.08\sqrt{\bar{E}_0\,\bar{w}_1}$ is lower than $\sigma_{p_2}\simeq1.11\sqrt{\bar{E}_0\,\bar{w}_1}$ and hence fracture is predicted at $x_2$. To further demonstrate the validity of this result, we perform the finite element analyses illustrated in Section \ref{subsubsct:2gm} with the same numerical and material parameters as in the analytical computations. The finite element results, given in Figure~\ref{fig:alpha_prof_diff}, are in agreement with the theoretical prediction of a crack at $x_2$.  

Further, Figure~\ref{fig:alpha_prof_diff} illustrates the normalized peak stresses corresponding to the two minima as functions of the regularization length of the phase-field model $\ell$. As $\ell$ decreases, the two normalized peak stresses decrease at a different rate, so that below a threshold value $\ell^* \simeq 0.056$ the failure location is shifted from $x_2$ to $x_1$, i.e. to the location predicted by the sharp-crack approach. 

Thus, in this example the non-local regularization inherent to the phase-field model may lead to different predictions on the failure location than the sharp-crack model, for which only the minimum value of the specific fracture energy counts. The predicted failure location of the phase-field model depends not only on the minimum value but also on the characteristics of its neighborhood (embodied in the parameter $r$). The amount of sampled neighborhood depends on the value of $\ell$, and the predicted failure location shifts to the one of the sharp-crack model as $\ell$ becomes sufficiently small. Interestingly, in this case it is the phase-field approach to be ill-posed for $\ell=\ell^*$. 

 \begin{figure}[H]
  \centering
  \hspace{-0.62em}
     \makebox[\textwidth][c]{\includegraphics{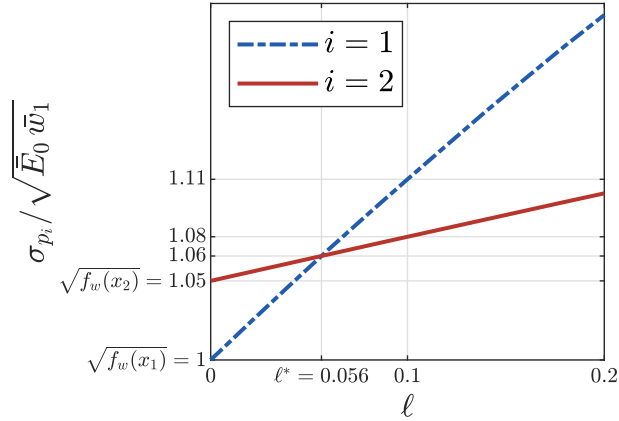}}
  \caption{Peak stresses for fracture at $x_1$ (blue curve) and at $x_2$ (red curve) vs.~internal length $\ell$ for the example with two different minima. The actual peak stress corresponds to the minimum among these two options and fracture occurs at the respective point. There is a value of length $\ell^*$ which marks a transition in the predicted failure location, from the one of the sharp-crack model $x_1$ ($\ell < \ell^*$) to the other one $x_2$ ($\ell > \ell^*$).}
  \label{fig:sigma_trend}
  \end{figure}
 
 
\section{Conclusions}
\label{sct:concl}
We investigated phase-field modeling of brittle fracture in a heterogeneous one-dimensional bar. We assumed the material properties to be minimum at the midpoint cross-section (taking these minimum values as reference material properties), and chose continuously and symmetrically increasing profiles of different shapes along the axis of the bar. Our main goal was to quantitatively assess how the heterogeneity in elastic and fracture material properties influences the observed tensile strength and fracture toughness of the bar, as obtained from the phase-field modeling approach.
\par The main findings can be summarized as follows:
\begin{itemize}
    \item The elastic limit stress for the heterogeneous bar is the same as for the bar made of the reference homogeneous material;
    \item The evolution problem for the heterogeneous bar does not admit a homogeneous solution in the damaging phase;
    \item Heterogeneous bars show a hardening branch after the elastic limit that leads to a peak stress larger than the elastic limit stress. The value of the peak stress is influenced by both elastic and fracture properties; for a given class and profile shape of heterogeneity, it only depends on the ratio between the internal length of the phase-field model and the length characterizing the speed of variation of the material properties (characteristic ratio);
    \item The fracture toughness for the heterogeneous bar is larger than for the bar made of the reference homogeneous material and is influenced by the fracture properties only; for a given profile shape of heterogeneity, it only depends on the characteristic ratio.
     \item The results obtained in the one-dimensional space are also valid for the bar embedded in the three-dimensional space. The only difference is that in the three-dimensional case the peak stress of the bar is also influenced by the Poisson's ratio, with a larger Poisson's ratio making the effect of heterogeneity more pronounced.
\end{itemize}
The observed effects of heterogeneity are direct consequences of the non-local nature of the phase-field model. This becomes evident through the comparison between sharp-crack and phase-field model predictions. Within the sharp-crack modeling framework, fracture in bars with heterogeneous specific fracture energy featuring multiple equal minima is an ill-posed problem and can only be addressed via stochastic relaxation. However, the same problem becomes well-posed with phase-field modeling if the heterogeneity profile features different values of the characteristic ratio at the equal minima. It is also shown that more complex cases of heterogeneity can be easily addressed by directly exploiting the findings in this study. In particular, the critical location among competing minima corresponding to "defects" of different sizes can be easily identified, which seems very relevant to the study of fracture in heterogeneous materials.


\FloatBarrier

\appendix


\section{Numerical solution of the localization problem}
\label{app:numericalDC}
Algorithm~\ref{alg:peak_stress} illustrates the determination of the peak stress, whereas Algorithm~\ref{alg:dp_plot} is used to plot the stress-displacement curve during the damage localization phase. In both, we collect the input lengths $\check{\delta}_t^{in}$ in a vector $\boldsymbol{\check{\delta}_{in}}$ and the corresponding output lengths $\check{\delta}_t^{out}$ for a given $\check{\sigma}_t$ in a vector $\boldsymbol{\check{\delta}_{out}}$. Eqs.~\ref{eq:dc_eta}-\ref{eq:central_slope_eta} provide the function $\boldsymbol{\check{\delta}_{out}}\gets\texttt{IVP}_{\sigma_t}(\boldsymbol{\check{\delta}_{in}})$ for each $\check{\sigma}_t$. In Algorithm~\ref{alg:peak_stress}, $q$ is the number of points along the piecewise linear interpolation of $(\boldsymbol{\check{\delta}_{in}},\boldsymbol{\check{\delta}_{out}})$ satisfying the condition $\check{\delta}_t^{out}=\check{\delta}_t^{in}$ (Figure~\ref{fig:D_function2}).
\RestyleAlgo{ruled}
\SetKwComment{Comment}{/* }{ */}

\begin{algorithm}
\caption{Determination of the peak stress}\label{alg:peak_stress}
\KwData{The function $\texttt{IVP}_{\check{\sigma}_t}$, the stress increment $\Delta\check{\sigma}=10^{-5}$, the half-support width increment $\Delta\check{\delta}_t=10^{-2}$ and the maximum half-support width $\check{\delta}_{max}=3$}
\KwResult{The dimensionless peak stress $\check{\sigma}_p$}
$\boldsymbol{\check{\delta}_{in}} \gets [1:\Delta\check{\delta}:\check{\delta}_{max}]$ \Comment*{Vector of the input lengths}
$\check{\sigma}_t \gets 1-\Delta \check{\sigma}$ \;
\Repeat{$q=0\textup{ }$}{
      $\check{\sigma}_t \gets \check{\sigma}_t+\Delta \check{\sigma}$ \;
      $\boldsymbol{\check{\delta}_{out}} \gets \texttt{IVP}_{\check{\sigma}_t}(\boldsymbol{\check{\delta}_{in}})$ \Comment*{Vector of the output lengths}
      $q\gets \texttt{Intersection}(\boldsymbol{\check{\delta}_{in}},\boldsymbol{\check{\delta}_{out}})$ \Comment*{Find the number $q$ of intersections between the piecewise linear interpolation of points $(\boldsymbol{\check{\delta}_{in}},\boldsymbol{\check{\delta}_{out}})$ and the straight line $\check{\delta}_t^{out}=\check{\delta}_t^{in}$}
    }
    $\check{\sigma}_p \gets \check{\sigma}_t$ 
\end{algorithm}


\SetKwComment{Comment}{/* }{ */}

\begin{algorithm}
\caption{Determination of the stress-displacement curve during damage localization}\label{alg:dp_plot}
\KwData{The function $\texttt{IVP}_{\check{\sigma}_t}$, the stress increment $\Delta\check{\sigma}=10^{-2}$, the half-support width increment $\Delta\check{\delta}=10^{-1}$, the maximum half-support width $\check{\delta}_{max}=3$, the ratios $\ell/2L$ and $\sqrt{\bar{w}_1/\bar{E}_0}$}
\KwResult{The displacement-stress sequence $(\boldsymbol{U}_t/2L,\boldsymbol{\check{\sigma}}_t)$ during the damage localization phase }
$\boldsymbol{\check{\delta}_{in}} \gets [1:\Delta\check{\delta}:\check{\delta}_{max}]$ \Comment*{Vector of the input lengths}
$\check{\sigma}_t \gets 0$ \;
$\boldsymbol{Q}\gets [\hspace{2mm}]$ \Comment*{Empty matrix}
\Repeat{$q=0$}{
      $\check{\sigma}_t \gets \check{\sigma}_t+\Delta \check{\sigma}$ \;
      $\boldsymbol{\check{\delta}_{out}} \gets \texttt{IVP}_{\check{\sigma}_t}(\boldsymbol{\check{\delta}_{in}})$ \Comment*{Vector of the output lengths}
      $\boldsymbol{\check{\delta}},q\gets \texttt{IntersectionPoints}(\boldsymbol{\check{\delta}_{in}},\boldsymbol{\check{\delta}_{out}})$ \Comment*{Find the points $(\boldsymbol{\check{\delta}},\boldsymbol{\check{\delta}})$ of intersection between the piecewise linear interpolation of points $(\boldsymbol{\check{\delta}_{in}},\boldsymbol{\check{\delta}_{out}})$ and the straight line $\check{\delta}_t^{out}=\check{\delta}_t^{in}$ and their multiplicity $q$}
      \newcommand{\forcond}{$\check{\delta}$ $\boldsymbol{in}$ $\boldsymbol{\check{\delta}}$}
      \For{\forcond}{
        $[\alpha^*_t,U_t,\check{\sigma}_t]\gets \texttt{IntegrateDC}(\check{\delta})$ \Comment*{Integrate the damage criterion using $\check{\delta}$ as input length of the semi-support and collect the maximum value of the damage, the applied displacement and the stress in a row vector}
        $\boldsymbol{Q}\gets \texttt{Append}\left([\alpha^*_t,U_t/2L,\check{\sigma}_t]\right)$ \Comment*{Append $[\alpha^*_t,U_t/2L,\check{\sigma}_t]$ to the matrix $\boldsymbol{Q}$ as new row}}}
    $\boldsymbol{Q_{\alpha}}\gets \texttt{SortRows}(\boldsymbol{Q})$ 
    \Comment*{Sort rows of $\boldsymbol{Q}$ in ascending order based on the value of the $\alpha^*_t$-column and save the new matrix as $\boldsymbol{Q_{\alpha}}$}
    $(\boldsymbol{U}_t/2L,\boldsymbol{\check{\sigma}}_t) \gets \texttt{PlotDS}(\boldsymbol{Q_{\alpha}})$ \Comment{Plot $(U_t/2L,\check{\sigma}_t)$ with the order in $\boldsymbol{Q_{\alpha}}$}
\end{algorithm}

\section{Lambert function}
  \label{app:Wfun}
\begin{figure}[H]
  \centering
     {\includegraphics{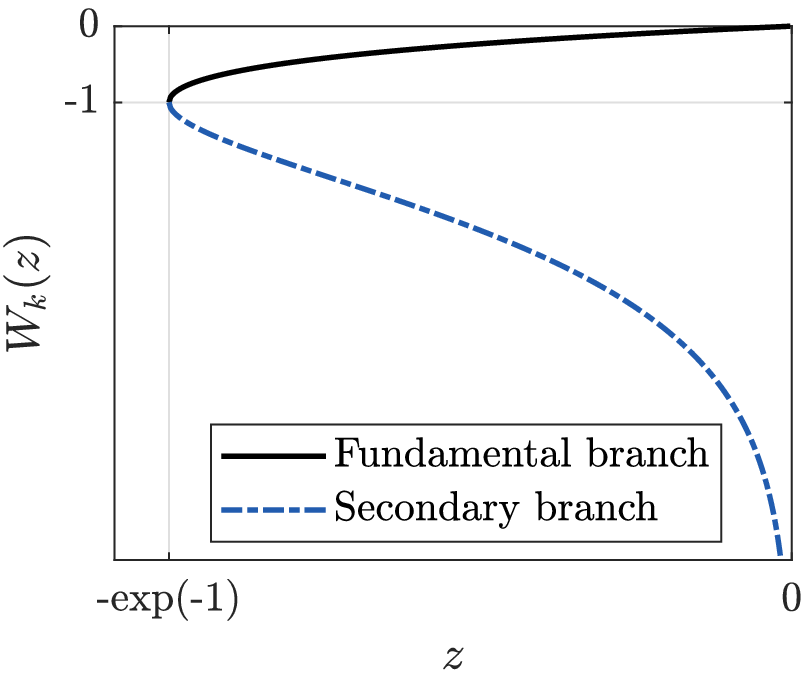}}
  \caption{Real branches of the Lambert function.}
  \label{fig:lambert}
  \end{figure}

The \emph{Lambert function} $W_k(z)$ is defined as \citep{corless1996lambertw}:
\begin{equation}
    \label{eq:lambert_definition}
 \tau=W_k(z), k\in \mathbb{Z}\hspace{5mm}\text{such that}\hspace{5mm}\tau\cdot \text{exp}(\tau)=z.
\end{equation}
This function has two real branches. The first one, termed \emph{fundamental branch}, is associated to $k=0$, while the second one, denoted as \emph{secondary branch}, is associated to $k=-1$.
The \emph{branch point} $\tau_0=-\text{exp}(-1)$ is the meeting point of the two real branches (Figure \ref{fig:lambert}).
  \section{Fracture toughness for $f_w$ with parabolic profile}
  \label{app:AT1_parabolic}
For $f_w$ with the parabolic profile shape,
the dimensionless damage criterion at $t=t_u$ within the half-support width reads
\begin{equation}
    2\,\alpha_u''(\check{x})+4\left(\frac{r^2\,\check{x}}{1+r^2\,\check{x}^2}\right)\alpha_u'(\check{x})=1\hspace{3mm}\text{in}\hspace{3mm}(-\check{\delta}_u,0).
\end{equation}
The analytical solution for the above differential equation depends on the unknown coefficients $c_1$ and $c_2$,
\begin{equation}
    \alpha_u(\check{x})=\frac{\check{x}^2}{12}+c_1\,\frac{\text{arctan}(r\check{x})}{r}+\frac{\text{log}(1+r^2\check{x}^2)}{6r^2} + c_2,
\end{equation}
\begin{equation}
    \alpha_u'(\check{x})=\frac{\check{x}}{6}+c_1\,\frac{1}{1+r^2\,\check{x}^2}+\frac{\check{x}}{3(1+r^2\,\check{x}^2)}.
\end{equation}
In order to find the two unknown coefficients as functions of the unknown half-support width we impose the two boundary conditions in Eq.~\ref{eq:bcs_delta}, obtaining
\begin{equation}
    c_1=\frac{1}{6}\,\check{\delta}_u\left(3+r^2\,\check{\delta}_u^2\right),
\end{equation}
\begin{equation}
    c_2=-\frac{\check{\delta}_u^2}{12}+\frac{\check{\delta}_u\,(3+r^2\,\check{\delta}_u^2)\,\text{arctan}(r\,\check{\delta}_u)}{6\,r}-\frac{\text{log}\left(1+r^2\,\check{\delta}_u^2\right)}{6\,r^2}.
\end{equation}
Numerically, we can find $\check{\delta}_u$ by enforcing the remaining boundary condition Eq.~\ref{eq:bcs2}.
\par
\begin{figure}[H]
  \centering
     {\includegraphics{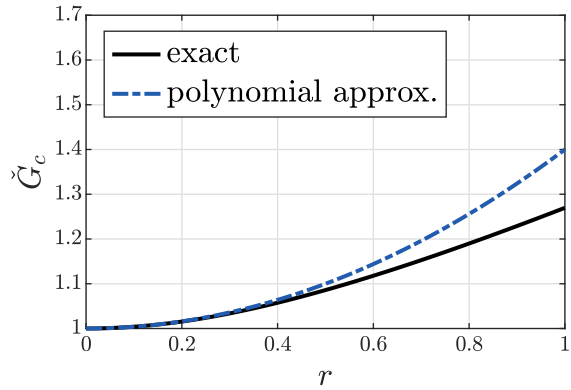}}
  \caption{Dimensionless fracture toughness vs.~characteristic ratio for $f_w$ with parabolic shape.}
  \label{fig:parabolic_Gc_appr}
  \end{figure}

 Eq.~\ref{eq:dimless_FT} yields the dimensionless fracture toughness
\begin{equation}
\label{parabolic_Gc}
\begin{aligned}
    \check{G}_c=\frac{1}{720\,r^3}\,& (
    -\check{\delta}_u\, r\,( 60+25\,\check{\delta}_u^2\,r^2+\\
    & +3\,\check{\delta}_u^4\,r^4)+15\left(1+\check{\delta}_u^2\,r^2\right)^2\left(4+\check{\delta}_u^2\,r^2\right)\arctan(\check{\delta}_u\,r)).
    \end{aligned}
\end{equation}
Combining the numerical solution for $\check{\delta}_u$ and Eq.~\ref{parabolic_Gc} we obtain the curve of the dimensionless fracture toughness vs.~the characteristic ratio in Figure~\ref{fig:parabolic_Gc_appr}.
\par In order to obtain a polynomial expansion of $\check{G}_c$ as a function of $r$ we perform a second order Taylor expansion of Eq.~\ref{eq:bcs2} about $r=0$.
This leads to the following fourth-order polynomial equation in $\check{\delta}_u$:
\begin{equation}
    \frac{1}{4}\,\check{\delta}_u^2+\frac{r^2}{12}\,\check{\delta}_u^4=1.
\end{equation}
Since $\check{\delta}_u$ must be real and non-negative for $r>0$, the solution is unique and equal to
\begin{equation}
\label{eq:delta_parabolic}
     \check{\delta}_u= \frac{\sqrt{-3 + \sqrt{3\,(3+16\,r^2)}}}{\sqrt{2}\,r}.   
\end{equation}
Combining Eqs.~\ref{parabolic_Gc} and \ref{eq:delta_parabolic} and taking again the second-order expansion about $r=0$, we obtain (Figure~\ref{fig:parabolic_Gc_appr})
\begin{equation}
    \label{eq:parabolic_app}
    \check{G}_c=1+\frac{2}{5}\,r^2+o(r^2).
\end{equation}
  
  
  \section{Fracture toughness for $f_w$ with exponential profile}
  \label{app:AT1_exponential}
\begin{figure}[H]
  \centering
     {\includegraphics{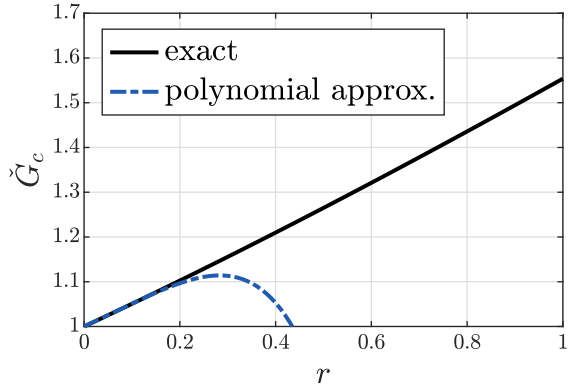}}
  \caption{Dimensionless fracture toughness vs.~characteristic ratio for $f_w$ with exponential shape.}
  \label{fig:exponential_Gc_approx}
  \end{figure}

For $f_w$ with the exponential profile shape,
the dimensionless damage criterion at $t=t_u$ within the half-support width reads
\begin{equation}
    2\,\alpha_u''(\check{x})-2\,r\,\alpha_u'(\check{x})=1\hspace{3mm}\text{in}\hspace{3mm}(-\check{\delta}_u,0).
\end{equation}
The analytical solution for the above differential equation depends on the unknown coefficients $c_1$ and $c_2$,
\begin{equation}
    \alpha_u(\check{x})=-c_1\,\frac{\text{exp}(r\,\check{x})}{r}-\frac{\check{x}}{2\,r}+c_2,
\end{equation}
\begin{equation}
    \alpha_u'(\check{x})=-\frac{1}{2r}-c_1\,\text{exp}(r\,\check{x}).
\end{equation}
The two unknowns are expressed as functions of the unknown half-support width through the two boundary conditions in Eq.~\ref{eq:bcs_delta}
\begin{equation}
    c_1=-\frac{\text{exp}(r\,\check{\delta}_u)}{2\,r}\hspace{3mm}\text{and}\hspace{3mm}c_2=-\frac{\check{\delta}_u}{2\,r}-\frac{1}{2\,r^2}.
\end{equation}
We can find $\check{\delta}_u$ by enforcing the remaining boundary condition in Eq.~\ref{eq:bcs2}, which yields
\begin{equation}
    \text{exp}(r\,\check{\delta}_u)=r\,\check{\delta}_u+1+2\,r^2.
\end{equation}
Through the substitution $\tau=-(r\,\check{\delta}_u+1+2\,r^2)$ and $z=-\text{exp}(-(1+2\,r^2))$, this equation becomes
\begin{equation}
z=\text{exp}(\tau)\cdot \tau,  
\end{equation}
hence (Section~\ref{subsct:FT})
\begin{equation}
\tau=W_{k}(z).
\end{equation}
Substituting backward and since $\check{\delta}_u>0$, the solution reads
\begin{equation}
\label{eq:delta_exp}
    \check{\delta}_u=-\frac{1}{r}[1+2\,r^2+W_{-1}(-\text{exp}(-(1+2\,r^2)))].
\end{equation}
Combining Eq.~\ref{eq:dimless_FT} and Eq.~\ref{eq:delta_exp}, the dimensionless fracture toughness is written in terms of the characteristic ratio (Figure~\ref{fig:exponential_Gc_approx})
\begin{equation}
    \begin{aligned}
    \check{G}_c=\frac{3}{16\,r^3}(
    1&-4\,r^2+\\
    &+W_{-1}(-\text{exp}(-1-2\,r^2))(2+
    W_{-1}(-\text{exp}(-1-2\,r^2)))
    ).
    \end{aligned}
\end{equation}
We simplify the expression for the dimensionless fracture toughness $\check{G}_c$ using the approximation of the Lambert function proposed by Veberi{\v c} \citep{veberivc2012lambert} and the Taylor expansion about $r=0$. Veberi{\v c}'s approximation is truncated at order 6 (\ref{app:lambert}) and the Taylor expansion at order 5 (Figure~\ref{fig:exponential_Gc_approx})
\begin{equation}
    \label{eq:exp_app}
    \check{G}_c\approx1+\frac{1}{2}\,r+\frac{2}{15}\,r^2-\frac{2}{135}\,r^3-\frac{5381}{1260}\,r^4-\frac{16147}{2835}\,r^5+o(r^5).
\end{equation}
  

\section{Non-linear equation for the half-support width for linear specific fracture energy}
\label{app:nonlin_delta}
Eq.~\ref{eq:bcs2} yields a non-linear equation in $\check{\delta}_u$ that reads
\begin{equation}
-r\,\check{\delta}_u\,(2+\check{\delta}_u\, r)+2\,(1+r\,\check{\delta}_u)^2\,\text{log}(1+r\,\check{\delta}_u)=8\,r^2.
\end{equation}
Through the substitutions $y=1+r\,\check{\delta}_u$ and $C=8\,r^2-1$, the equation can be rewritten as
\begin{equation}
y^2\left(2\,\text{log}(y)-1\right)=C,
\end{equation}
that can be rearranged as follows:
\begin{equation}
\frac{C}{\text{exp}(1)}=\frac{C}{y^2}\,\text{exp}\left(\frac{C}{y^2}\right).
\end{equation}
Proceeding with the further substitutions $z=\frac{C}{\text{exp}(1)}$ and $\tau=C/y^2$ we retrieve Eq.~\ref{eq:delta_lambert}.


\section{Veberi{\v c} approximation of the Lambert function}
\label{app:lambert}

Veberi{\v c} \citep{veberivc2012lambert} proposes an approximation of the two branches of the Lambert function about the branch point $\tau_0$, based on the Taylor expansion of the inverse of the Lambert function
\begin{equation}
  W_{0,-1}(z)\approx\sum_{i=0}^n m_i\cdot b^i_{\pm}(z)\hspace{3mm}\text{with}\hspace{3mm} b_{\pm}(z)=\pm\sqrt{\left(2\left(1+\text{exp}(1)\cdot z\right)\right)},
\end{equation}
where $+$ is referred to $k=0$ and $-$ is associated to $k=-1$. The first coefficients $m_i$ are reported in Table \ref{Tab:veberic}. 
\begin{table}[H]\centering
  \begin{tabular}{c c c c c c c c} \toprule	
		\multirow{2}{*}{$i$} & 
		\multirow{2}{*}{$0$} &
		\multirow{2}{*}{$1$} &
		\multirow{2}{*}{$2$} &
		\multirow{2}{*}{$3$} &
		\multirow{2}{*}{$4$} &
		\multirow{2}{*}{$5$} &
		\multirow{2}{*}{$6$} \\
		\\\toprule

  $m_i$  	&	$-1$	    &	$1$ &  $-\frac{1}{3}$ &$\frac{11}{72}$ &$-\frac{43}{540}$ &$-\frac{769}{17280}$  &$-\frac{221}{8505}$ 	\\		\bottomrule
\end{tabular}
\caption{Coefficients for Veberi{\v c}'s approximation.}
  \label{Tab:veberic}
  \end{table}
  
  Here we present only 7 coefficients but it is possible to compute an arbitrary amount of coefficients following the procedure outlined in \citep{veberivc2012lambert}.

\section{Material parameters}
\label{app:mat_par}
The undamaged elasticity tensor depends on the two undamaged Lamé parameters  $\lambda_0(\boldsymbol{x})$ and $\mu_0(\boldsymbol{x})$ as
\begin{equation}
        \mathbb{C}_{0}(\boldsymbol{x})=\lambda_0(\boldsymbol{x}) \boldsymbol{I}\otimes \boldsymbol{I}+2\mu_0(\boldsymbol{x}) \mathbb{I}^s,
\end{equation}
 hence,
\begin{equation}
    \label{eq:het_lam}
    \lambda_0(\boldsymbol{x})=\bar{\lambda}_0\cdot f_E(x)\hspace{3mm}\text{and}\hspace{3mm}\mu_0(\boldsymbol{x})=\bar{\mu}_0\cdot f_E(x),
\end{equation}
where $\bar{\lambda}_0$ and $\bar{\mu}_0$ are independent of $\boldsymbol{x}$. From the Lamé parameters, the undamaged elastic modulus $E_0$ and the Poisson's ratio $\nu$ are derived as
\begin{equation}
    \label{eq:eng_lam}
    E_0(\boldsymbol{x})=\frac{\mu_0(\boldsymbol{x})(3\lambda_0(\boldsymbol{x})+2\mu_0(\boldsymbol{x}))}{\lambda_0(\boldsymbol{x})+\mu_0(\boldsymbol{x})}\hspace{3mm}\text{and}\hspace{3mm}\nu(\boldsymbol{x})=\frac{\lambda_0(\boldsymbol{x})}{2(\lambda_0(\boldsymbol{x})+\mu_0(\boldsymbol{x}))}.
\end{equation}
Therefore,
\begin{equation}
    E_0(\boldsymbol{x})=\bar{E}_0\cdot f_E(x)
    \hspace{3mm}\text{and}\hspace{3mm}
    \nu(\boldsymbol{x})=\bar{\nu}
\end{equation}
whith $\bar{E}_0=\frac{\bar{\mu}_0\,(3\,\bar{\lambda}_0+2\,\bar{\mu}_0)}{\bar{\lambda}_0+\bar{\mu}_0}$ and $\bar{\nu}=\frac{\bar{\lambda}_0}{2\,(\bar{\lambda}_0+\bar{\mu}_0)}$.

\section*{Acknowledgements}
\par This research has received funding from the European Union’s Horizon 2020 research and
innovation programme under the Marie Skłodowska-Curie grant agreement No. 861061 – NEWFRAC Project. The authors would also like to acknowledge Prof. Corrado Maurini for helpful discussion. 

\bibliographystyle{elsarticle-num}
\bibliography{ms} 

\begin{thebibliography}{10}
\expandafter\ifx\csname url\endcsname\relax
  \def\url#1{\texttt{#1}}\fi
\expandafter\ifx\csname urlprefix\endcsname\relax\def\urlprefix{URL }\fi
\expandafter\ifx\csname href\endcsname\relax
  \def\href#1#2{#2} \def\path#1{#1}\fi

\bibitem{bourdin2000numerical}
B.~Bourdin, G.~A. Francfort, J.-J. Marigo, Numerical experiments in revisited
  brittle fracture, Journal of the Mechanics and Physics of Solids 48~(4)
  (2000) 797--826.

\bibitem{francfort1998revisiting}
G.~A. Francfort, J.-J. Marigo, Revisiting brittle fracture as an energy
  minimization problem, Journal of the Mechanics and Physics of Solids 46~(8)
  (1998) 1319--1342.

\bibitem{pham2011gradient}
K.~Pham, H.~Amor, J.-J. Marigo, C.~Maurini, Gradient damage models and their
  use to approximate brittle fracture, International Journal of Damage
  Mechanics 20~(4) (2011) 618--652.

\bibitem{ambati2015review}
M.~Ambati, T.~Gerasimov, L.~De~Lorenzis, A review on phase-field models of
  brittle fracture and a new fast hybrid formulation, Computational Mechanics
  55~(2) (2015) 383--405.

\bibitem{natarajan2019phase}
S.~Natarajan, R.~K. Annabattula, E.~Mart{\'\i}nez-Pa{\~n}eda, et~al., Phase
  field modelling of crack propagation in functionally graded materials,
  Composites Part B: Engineering 169 (2019) 239--248.

\bibitem{kumar2021phase}
P.~A.~V. Kumar, A.~Dean, J.~Reinoso, P.~Lenarda, M.~Paggi, Phase field modeling
  of fracture in functionally graded materials: $\gamma$-convergence and
  mechanical insight on the effect of grading, Thin-Walled Structures 159
  (2021) 107234.

\bibitem{hossain2014effective}
M.~Hossain, C.-J. Hsueh, B.~Bourdin, K.~Bhattacharya, Effective toughness of
  heterogeneous media, Journal of the Mechanics and Physics of Solids 71 (2014)
  15--32.

\bibitem{shen2019novel}
R.~Shen, H.~Waisman, Z.~Yosibash, G.~Dahan, A novel phase field method for
  modeling the fracture of long bones, International Journal for Numerical
  Methods in Biomedical Engineering 35~(8) (2019) e3211.

\bibitem{pham2013onset}
K.~Pham, J.-J. Marigo, From the onset of damage to rupture: construction of
  responses with damage localization for a general class of gradient damage
  models, Continuum Mechanics and Thermodynamics 25~(2) (2013) 147--171.

\bibitem{pham2010approche}
K.~Pham, J.-J. Marigo, Approche variationnelle de l'endommagement: I. les
  concepts fondamentaux, Comptes Rendus M{\'e}canique 338~(4) (2010) 191--198.

\bibitem{pham2010approcheb}
K.~Pham, J.-J. Marigo, Approche variationnelle de l'endommagement: {II}. les
  mod{\`e}les {\`a} gradient, Comptes Rendus M{\'e}canique 338~(4) (2010)
  199--206.

\bibitem{gerasimov2019penalization}
T.~Gerasimov, L.~De~Lorenzis, On penalization in variational phase-field models
  of brittle fracture, Computer Methods in Applied Mechanics and Engineering
  354 (2019) 990--1026.

\bibitem{shampine1997matlab}
L.~F. Shampine, M.~W. Reichelt, The matlab ode suite, SIAM journal on
  Scientific Computing 18~(1) (1997) 1--22.

\bibitem{miehe2010thermodynamically}
C.~Miehe, F.~Welschinger, M.~Hofacker, Thermodynamically consistent phase-field
  models of fracture: Variational principles and multi-field fe
  implementations, International Journal for Numerical Methods in Engineering
  83~(10) (2010) 1273--1311.

\bibitem{veberivc2012lambert}
D.~Veberi{\v{c}}, Lambert w function for applications in physics, Computer
  Physics Communications 183~(12) (2012) 2622--2628.

\bibitem{griphfit2022}
{D-MAVT IMES Computational Mechanics Group, ETH Z\"urich}, {GRIPHFiTH - GRoup
  Implementation for PHase-field Fracture THeory},
  \url{https://www.doi.org/10.5905/ethz-1007-523}, visited on 2022-07-15
  (2022).

\bibitem{bourdin2007numerical}
B.~Bourdin, Numerical implementation of the variational formulation for
  quasi-static brittle fracture, Interfaces and free boundaries 9~(3) (2007)
  411--430.

\bibitem{currey1988effect}
J.~D. Currey, The effect of porosity and mineral content on the young's modulus
  of elasticity of compact bone, Journal of Biomechanics 21~(2) (1988)
  131--139.

\bibitem{katz2019scanner}
Y.~Katz, G.~Dahan, J.~Sosna, I.~Shelef, E.~Cherniavsky, Z.~Yosibash, Scanner
  influence on the mechanical response of qct-based finite element analysis of
  long bones, Journal of biomechanics 86 (2019) 149--159.

\bibitem{francfort1999cracks}
G.~Francfort, J.~Marigo, Cracks in fracture mechanics: A time indexed family of
  energy minimizers, in: IUTAM symposium on variations of domain and
  free-boundary problems in solid mechanics, Springer, 1999, pp. 197--202.

\bibitem{gerasimov2020stochastic}
T.~Gerasimov, U.~R{\"o}mer, J.~Vond{\v{r}}ejc, H.~G. Matthies, L.~De~Lorenzis,
  Stochastic phase-field modeling of brittle fracture: computing multiple crack
  patterns and their probabilities, Computer Methods in Applied Mechanics and
  Engineering 372 (2020) 113353.

\bibitem{corless1996lambertw}
R.~M. Corless, G.~H. Gonnet, D.~E. Hare, D.~J. Jeffrey, D.~E. Knuth, On the
  lambert w function, Advances in Computational Mathematics 5~(1) (1996)
  329--359.

\end{thebibliography}

\end{document}